\documentclass[a4paper]{aa}
\usepackage[varg]{txfonts}
\pdfoutput=1

\usepackage{hyperref}
\hypersetup{colorlinks=true,linkcolor=blue,citecolor=blue,filecolor=blue,urlcolor=blue}

\usepackage{graphicx}
\graphicspath{{images/}}

\usepackage{bm}

\usepackage{booktabs,dcolumn}
\newcolumntype{d}[1]{D{.}{.}{#1}}

\newcommand{\der}{\mathrm{d}}
\newcommand{\tcr}{t_\mathrm{cr}}
\newcommand{\tcc}{t_\mathrm{cc}}
\newcommand{\tbin}{t_\mathrm{bin}}
\newcommand{\Ebin}{E_\mathrm{bin}}
\newcommand{\Elim}{E_\mathrm{lim}}
\newcommand{\rstar}{r_\star}
\newcommand{\rc}{r_\mathrm{c}}
\newcommand{\rhostar}{\rho_\star}
\newcommand{\rhoc}{\rho_\mathrm{c}}
\newcommand{\mstar}{m_\star}
\newcommand{\mc}{m_\mathrm{c}}

\newcommand{\tot}{\mathrm{tot}}
\newcommand{\rlmin}{r_\mathrm{m}}
\newcommand{\rlmini}{r_{\mathrm{m},i}}
\newcommand{\tmin}{t_\mathrm{m}}
\newcommand{\tmini}{t_{\mathrm{m},i}}

\newcommand{\I}{\mathrm{I}}
\newcommand{\II}{\mathrm{I\!I}}
\newcommand{\III}{\mathrm{I\!I\!I}}
\newcommand{\IV}{\mathrm{I\!V}}

\newcommand{\emod}{\texttt{e-10k}}
\newcommand{\emodi}{\texttt{e-50k}}
\newcommand{\mmod}{\texttt{m-20k}}
\newcommand{\mmodi}{\texttt{m-100k}}
\newcommand{\mmodii}{\texttt{n-100k}}

\defcitealias{gen_evol}{Storn--Price}
\defcitealias{savitzky_golay}{Savitzky--Golay}

\begin{document}

\title{The hunt for self-similar core collapse}
\subtitle{}

% persuade the template not to print email footnotes with \dagger (4) and \ddagger (5) instead of \star (0) and \star\star (1)
\setcounter{footnote}{4}

\author{Václav Pavlík\inst{\ref{auuk},\ref{obs},}\thanks{\email{pavlik@sirrah.troja.mff.cuni.cz.}}
\and Ladislav Šubr\inst{\ref{auuk}}}

\institute{Astronomical Institute of Charles University, Prague, Czech Republic\label{auuk}
\and Observatory and Planetarium of Prague, Prague, Czech Republic\label{obs}}

\titlerunning{The hunt for self-similar core collapse}
\authorrunning{Pavlík \& Šubr}

\date{Received: 16 July 2018 / Accepted: 13 August 2018}

\abstract
{Core collapse is a prominent evolutionary stage of self-gravitating systems. In an idealised collisionless approximation, the region around the cluster core evolves in a self-similar way prior to the core collapse. Thus, its radial density profile outside the core can be described by a power law, $\rho \propto r^{-\alpha}$.}
{We aim to find the characteristics of core collapse in $N$-body models. In such systems, a complete collapse is prevented by transferring the binding energy of the cluster to binary stars. The contraction is, therefore, more difficult to identify.}
{We developed a method that identifies the core collapse in $N$-body models of star clusters based on the assumption of their homologous evolution.}
{We analysed different models (equal- and multi-mass), most of which exhibit patterns of homologous evolution, yet with significantly different values of $\alpha$ : the equal-mass models have $\alpha \approx 2.3$, which agrees with theoretical expectations, the multi-mass models have $\alpha \approx 1.5$ (yet with larger uncertainty). Furthermore, most models usually show sequences of separated homologous collapses with similar properties.
Finally, we investigated a correlation between the time of core collapse and the time of formation of the first hard binary star. The binding energy of such a binary usually depends on the depth of the collapse in which it forms, for example\ from $100\,kT$ to $10^4\,kT$ in the smallest equal-mass to the largest multi-mass model, respectively. However, not all major hardenings of binaries happened during the core collapse. In the multi-mass models, we see large transfers of binding energy of $\sim 10^4\,kT$ to binaries that occur on the crossing timescale and outside of the periods of the homologous collapses.}
{}

\keywords{methods: numerical -- galaxies: clusters: general -- stars: binaries: general}

\maketitle

\section{Introduction}
\label{sec:intro}

Two-body relaxation states that any star moving through a field of stars is decelerated by the force of dynamical friction \citep{chandrasekhar}, which is proportional to the mass of the moving star and inversely proportional to its velocity squared.
This force is also responsible for mass segregation in multi-mass systems.
Since the heat capacity of a self-gravitating system in virial equilibrium is negative \citep[e.g.][]{lb_w,binney_tremaine}, the central region of a star cluster should contract over time.
Consequently, the central parts relax more quickly than the halo and the velocity distribution in the centre is almost Maxwellian \citep{larson}.
Within the thermodynamic framework, the core is supposed to collapse, reaching infinite density and kinetic temperature in a finite time (also known as the gravothermal catastrophe). Unlike continuum models, in $N$-body models (and real star clusters) this sequence is prevented by the presence of existing or newly formed binary stars in the core, whose ability to efficiently expel other stars via three-body interactions cools the core \citep[e.g.][]{aarseth1972,hut,fujii_pz,oleary}. The cluster core gradually shrinks towards the collapse and then expands rapidly (so called core bounce). Thus, the event of core collapse may be indirectly observed but its exact time is no longer well defined.

\cite{lb_e} showed that the evolution of a spherically symmetric collisionless system prior to core collapse should be self-similar (homologous), that is\ its density evolves with respect to the radius and time according to the \hbox{scaling relation}
\begin{equation}
  \label{eq:density}
  \rho(r, t) = \rhoc(t) \, \rhostar(\rstar) \,.
\end{equation}
Here, $\rhoc$ is the core density, $\rhostar$ is a dimensionless structure function, and the radius is described using an enclosed mass $m$ and a scaling factor $\rstar$,
\begin{equation}
  \label{eq:radius}
  r(m, t) = \rc(t) \, \rstar\!\left(\frac{m}{\mc}\right) \,,
\end{equation}
where $\rc$ stands for the core radius and $\mc \propto \rhoc \rc^3$ is the core mass.
The homologous solution implies that the internal structural scaling has an exponent $\alpha$ that remains temporally invariant. As it must also satisfy smoothness conditions for $\rhostar(\rstar)$ and normalisation, generally $\alpha = \mathrm{const.}$ (\citealt{lb_e}, also e.g.\ \citealt{penston}). The core radius then depends on the core density as $\rhoc \propto \rc^{-\alpha}$ and the temporal evolution of the core radius before the time of core collapse, $\tcc$, is
\begin{equation}
  \label{eq:self_similar}
  \rc(t) \propto (\tcc - t)^{\frac{2}{6 - \alpha}} \,.
\end{equation}

\begin{figure}
  \includegraphics[width=\linewidth]{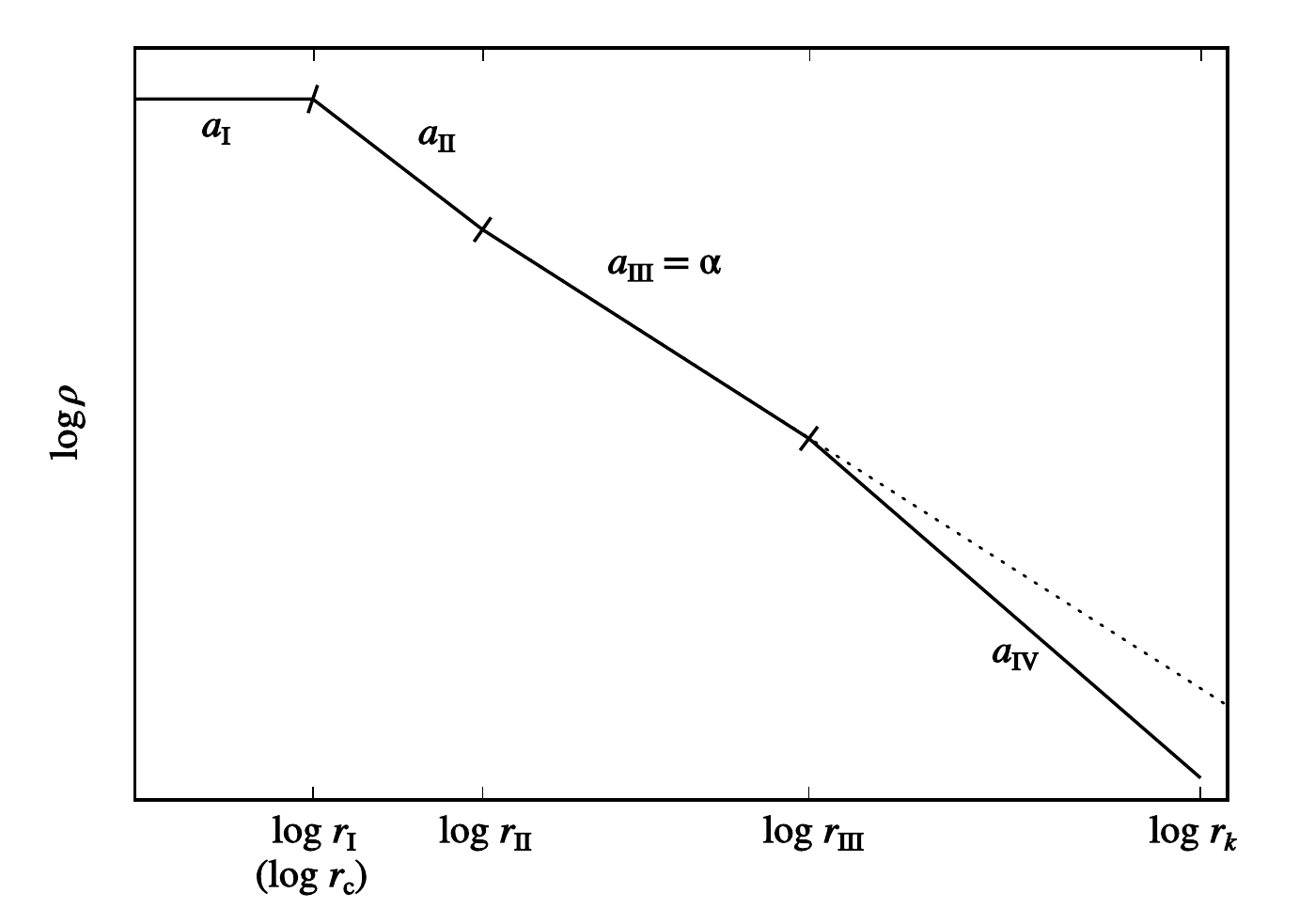}
  \caption{Schematic plot of the radial density profile in an $N$-body cluster during the core collapse with four different values of the logarithmic density gradient, $a_{\I-\IV}$\,. The dotted line has a slope equal to $\alpha$ and shows the asymptotic solution of \citet{lb_e}. The break radius $r_\I$ is identified with the core radius, $r_\III$ roughly corresponds to the half-mass radius, and $r_k$ is the cluster radius.
  The notation is the same as in Eq. \eqref{eq:broken} that we used for the fitting of our numerical models.}
  \label{fig:schema}
\end{figure}

According to \cite{lb_e}, the radial density profile may be approximated by a double-broken power-law function with the logarithmic density gradient defined as
\begin{equation}
  \label{eq:alpha}
  a \equiv -\frac{\der \log{\rho}}{\der \log{r}} \,
,\end{equation}
which is equal to zero in the cluster core and reaches $\alpha$ asymptotically. In an intermediate region above $\rc$, the logarithmic density gradient has to be larger than $\alpha$ to compensate for the missing mass in the core (see the slopes $a_\I$, $a_\II$ , and $\alpha$ in Fig.~\ref{fig:schema}).
\citet{lb_e} found $\alpha \approx 2.208$. Further works based on either isotropic \citep{cohn} or anisotropic models \citep{takahashi} in a Fokker--Planck approximation led to a slightly different value, $\alpha \approx 2.23$.

In $N$-body star clusters, the self-similar solution is not infinite. In this case, distant parts of the halo tend to a Maxwellian distribution of velocities on the relaxation timescale. This changes the logarithmic density gradient in the halo to $a = 3.5$ \citep{spitzer_hart}. Hence, we expect the cluster's radial density profile to be approximated by a triple-broken power law as indicated by a solid line in Fig.~\ref{fig:schema}.

Radial density profiles of numerical $N$-body models \citep[e.g.][]{giersz_heggie,makino,core_collapse} show a good agreement with the expectations above. In this paper we go beyond those works in several aspects: (i) computational resources available nowadays enable us to integrate and analyse hundreds of realisations of models consisting of tens of thousands of particles; (ii) besides equal-mass models, we evaluate models with a Salpeter mass function and; (iii) we not only analyse the density profile of the cluster at the time of core collapse but also test the hypothesis of a homologous evolution in time which, among other things, allows us to formulate a new method for determining the time of core collapse.

\section{Numerical models}

The numerical models that we are working with may be scaled arbitrarily, so we express all numbers in this paper in Hénon units (unless we specify otherwise) that are used in $N$-body integrations, that is,\ $G = M_\tot = R_\mathrm{virial} = 1$\,. The total energy is $E_\tot = -\frac{1}{4}$ and the crossing time is $\tcr = 2\sqrt{2}$ in those units.

We studied several $N$-body models of star clusters: two equal-mass (\emod, \emodi) and three multi-mass (\mmod, \mmodi, \mmodii) with a \citet{salpeter} slope of $-2.35$ for the initial mass function in the mass range listed in Table~\ref{tab:models}. We generated initial conditions of all models using the \texttt{plumix} code \citep{subr,plumix} without initial mass segregation, that is\ with a \citet{plummer} distribution function. The dynamical evolution was followed using the collisional \texttt{nbody6} code \citep{aarseth}.

Being concerned only with dynamical effects, our star clusters were modelled as isolated systems, and we did not consider any external potential or tidal forces from the Galaxy. Stars were represented by point masses and we did not include any primordial binaries or stellar evolution. Collisions of stars were disabled and escaping stars were not removed from any simulation. All models are pure $N$-body and neither gas nor dust components were included.
In other words, this work is an exploration of the dynamics of isolated, ideal self-gravitating systems.

\begin{table}
  \centering
  \caption{Parameters and notation of the numerical models analysed in this paper: \# is the number of realisations, $N_\tot$ is the total number of particles, IMF is either the slope of the initial mass function or expressed as equal masses, and $m_\star$ represents the mass (or range of masses) of individual stars.}
  \label{tab:models}
  \begin{tabular}{ccccc}
    \hline
    name                  & \#          & $N_\tot$ & IMF             & $m_\star$\\
    \hline
    \emod & 85  & \,\ 10\,000  & eq.\,m.    & $1.0{\times}10^{-4}$ \\
    \emodi & 10  & \,\ 50\,000  & eq.\,m.    & $2.0{\times}10^{-5}$ \\
    \mmod & 100  & \,\ 20\,000  & $-2.35$ & $1.5{\times}10^{-5} - 3.8{\times}10^{-3}$ \\
    \mmodi & 10  & 100\,000 & $-2.35$ & $3.0{\times}10^{-6} - 1.5{\times}10^{-3}$ \\
    \mmodii & 10  & 100\,000 & $-2.35$ & $3.8{\times}10^{-6} - 9.5{\times}10^{-5}$ \\
    \hline
  \end{tabular}
\end{table}

\section{Methods}

\subsection{Time of core collapse}
\label{sec:method_cc}

One definition of the time of core collapse is the moment of the minimum core radius. However, when the cluster core shrinks, its binding energy rises and all stars that are moving inward or outward have increasingly greater speeds. The core becomes unstable and subject to random pulsations \citep{makino_oscillations}, and the definition of the core is not obvious. Consequently, the time of its minimum radius is not clearly defined.
An alternative method, which we will discuss later, is related to the formation of the first hard binary system \citep{fujii_pz}. In this section, we present a third method that is based on analytic predictions of the self-similar evolution of a star cluster before the core collapse.

The evolution of star clusters is often described by means of the Lagrangian radii (e.g.\ see Figs.~\ref{fig:lagr_s10k} to~\ref{fig:lagr_m100k}), which are defined as the radii of concentric spheres containing fixed mass fractions. This definition is obviously dependent on the choice of the origin of the coordinate system because at each time step, stars that occupy these spherical regions are reordered by their current radial distance from the origin. In our models, we calculate the Lagrangian radii from the cluster's density centre provided by \texttt{nbody6} \citep[i.e.\ a numerically improved method from][which is a well established and commonly used approach]{casertano_hut}.

In a homologous solution, the evolution of the Lagrangian radii is given by Eq. \eqref{eq:radius}. At the moment of their minima, the Lagrangian radii are related to $\rc$ by a single constant of proportionality. The minimum of every Lagrangian radius curve is given by
\begin{equation}
  \label{eq:lr_condition}
  \left(\frac{\partial r}{\partial t}\right)_{\!m} = 0 \,.
\end{equation}
After substituting $r$ in \eqref{eq:lr_condition} from Eq. \eqref{eq:radius}, we get
\begin{equation}
  \label{eq:der_r}
  \frac{\mc}{\dot{\mc}}\frac{\dot{r}_\mathrm{c}}{\rc} = \frac{\mstar}{\rstar}\frac{\partial \rstar}{\partial \mstar} \,,
\end{equation}
where a dot represents $\frac{\partial}{\partial t}$ and, using homology, $\mstar = m / \mc(t)$\,. Following the same argument as \citet{lb_e} that each side of this equation depends on a different parameter ($t$ and $\mstar$, respectively), both sides must be constant with respect to these parameters. Hence, the scaling functions $\mstar$ and $\rstar$ have the same value for all minima of the Lagrangian radii. This implies that the radius representing the minimum (hereafter $\rlmin$) is a fixed multiple of $\rc$ at the time of its minimum (denoted as $\tmin$). Together with Eq. \eqref{eq:self_similar}, we have
\begin{equation}
  \label{eq:self_similar_rl}
  \rlmin \propto \big( \tcc - \tmin \big)^{\frac{2}{6 - \alpha}} \,.
\end{equation}
Consequently, parameters of the homologous solution, $\tcc$ and $\alpha$, are obtainable from fitting function \eqref{eq:self_similar_rl} to the sequence of minima of the Lagrangian radii in numerical models.

\subsubsection{Data processing and fitting}

As mentioned above, the cluster core pulsates with unpredictable frequency and amplitude while collapsing. Fluctuations of the unstable core also propagate, by definition, in the Lagrangian radii. They are most visible for low mass fractions, that is,\ inner Lagrangian radii. These fluctuations arise from random motions of stars in the cluster core. We are, however, interested in general trends that are driven by the cluster thermodynamics. In order to reveal them and, in particular, to find the corresponding minima of the Lagrangian radii for Eq. \eqref{eq:self_similar_rl}, we reduced the fluctuations by smoothing each Lagrangian radius time-wise. We tried a moving average and an algorithm from  \citet{savitzky_golay}, which is a method with a higher order polynomial. The latter proved better in reducing the fluctuations while preserving the global evolution of the Lagrangian radii, so we will focus only on that one. 

We found the minima of $n$ smoothed Lagrangian radii that provide us with a set of points
\begin{equation}
  \label{eq:lr_minima}
  \left\lbrace \big( \tmini,\rlmini \big) : i \in \mathbb{N}^* \wedge i \leq n \right\rbrace \,.
\end{equation}
Assuming that these points of minima satisfy the homologous solution described by Eq. \eqref{eq:self_similar_rl}, we may find $\tcc$ by fitting them with function
\begin{equation}
  \label{eq:fit}
  \rlmin(\tmin) = A\ (\tcc - \tmin)^{\frac{2}{6 - \alpha}}
\end{equation}
via the parameters $\tcc$, $\alpha$ , and a proportionality constant $A$. This function may only be used for $\tmin \leq \tcc$ , which makes $\tcc$ not only a fitting parameter but also an upper bound for $\tmin$. It is evident that, for example Eq. \eqref{eq:fit} cannot be used with a fitting method that requires fixed limits; the solution in such cases is to introduce a Heaviside function
\begin{equation}
  H(x) =
  \begin{cases}
    1 \,, & \text{for } x \geq 0 \\
    0 \,, & \text{otherwise}
  \end{cases}
\end{equation}
and then redefine Eq. \eqref{eq:fit} to
\begin{equation}
  \label{eq:fit_heaviside}
  \rlmin(\tmin) = A\ H(\tcc - \tmin)\ (\tcc - \tmin)^{\frac{2}{6 - \alpha}} \,,
\end{equation}
where $\tmin$ is formally no longer restricted by the choice of bounds.

The best fit is determined by the lowest value of a sum of the squared absolute deviations in radii, which we call $s^2_r$, defined as
\begin{equation}
  s^2_r = \sum_{i=1}^{n}{\big[ \rlmini - \rlmin(\tmini) \big]^2} \,.
\end{equation}
As a minimising method, we applied a genetic algorithm by \citet{gen_evol}.\footnote{The algorithm is implemented in the \texttt{python} library \texttt{scipy.optimize} as \texttt{differential\_evolution()}.}\!

\subsubsection{Limits}
\label{sec:limits}

The \citetalias{gen_evol} minimisation algorithm requires fixed lower and upper limits for each parameter to generate an initial random sample of seeds in the parameter space.
The range of possible values of $\alpha$ must satisfy the condition $\alpha \neq 6$ that comes from the exponent $\frac{2}{6 - \alpha}$ in Eq. \eqref{eq:self_similar}. According to the analytic solution, $\alpha$ is supposed to reach a value between $2$ and $2.5$ \citep[cf.][]{lb_e}, hence we set the upper limit to $\alpha < 6$. Assuming that the density is a decreasing function of distance from the centre, we set the lower limit to $\alpha \geq 0$.

The points of minima \eqref{eq:lr_minima} are supposed to represent a decreasing power-law function, so we do not expect the time of core collapse to happen sooner than the earliest minimum of the Lagrangian radii we fitted. Therefore, the lower limit of $\tcc$ was set to $\min{(\tmini)}$\,. The upper limit is a value that is reasonably large, so it will not affect the results, that is,\ $2\max{(\tmini)} - \min{(\tmini)}\,.$
The range of the proportionality constant $A$ was set wide enough not to restrict the fit,
for example\ from $-2$ to $2$, whereas the typical values of $A$ were around $0.008$ in model \emod\ and around $0.04$ in model \mmod.

\subsection{Density profile at core collapse}
\label{sec:method_rho}

To evaluate the radial density profile of a star cluster during the core collapse, we linearly interpolated each Lagrangian radius to the determined $\tcc$ from the nearest values around that time. Hereafter, we denote the $i$-th Lagrangian radius at core collapse by $r_i$. So, at $\tcc$, we have a set of Lagrangian radii
\begin{equation}
  \label{eq:lr_Tcc}
  \left\lbrace r_i : i \in \mathbb{N}_0 \wedge i \leq k \right\rbrace \,,
\end{equation}
where $r_1$ corresponds to the lowest mass fraction, $r_k$ is the cluster radius (defined here by the Lagrangian radius of $99\,\%$ of mass), and $r_0 \equiv 0$. The mean density contained in the spherical shell between every pair of consecutive Lagrangian radii (assuming spherical symmetry) is given by
\begin{equation}
  \label{eq:shell}
  \rho_i = \frac{3}{4 \pi} \frac{m_{i+1} - m_{i}}{r_{i+1}^3 - r_{i}^3} \,, \text{ for } i \in \left\lbrace 0, 1, \dots, k - 1 \right\rbrace \,,
\end{equation}
where $m_{i}$ is the mass percentage (i.e.\ the mass in Hénon units) contained in a sphere of radius $r_i$. By definition, $m_0 = m(r_0) \equiv 0$. Finally, we have a set of points
% \begin{equation}
%   \label{eq:lr_rho}
%   \left\lbrace \big( r_i, \rho_i \big) : i \in \mathbb{N}_0 \wedge i \leq k \right\rbrace
% \end{equation}
$\left\lbrace ( r_i, \rho_i ) \right\rbrace$ representing the radial density profile of a star cluster at the time of core collapse.

\subsubsection{Fitting}

We fit a triple-broken power-law function (see Fig.~\ref{fig:schema})
\begin{equation}
  \label{eq:broken}
  \rho(r) =
  \begin{cases}
    b_\I r^{-a_\I} \,, \text{ for } 0 \leq r \leq r_\I \\
    b_\II r^{-a_\II} \,, \text{ for } r_\I \leq r \leq r_\II \\
    b_\III r^{-a_\III} \,, \text{ for } r_\II \leq r \leq r_\III \\
    b_\IV r^{-a_\IV} \,, \text{ for } r_\III \leq r \leq r_k
  \end{cases}
\end{equation}
to the set $\left\lbrace ( r_i, \rho_i ) \right\rbrace$.
The variables $r_\I$, $r_\II$ , and $r_\III$ denote the radii where this power-law function breaks, and $r_k$ is the cluster radius (in our case the Lagrangian radius of 99\,\% of mass). Following the discussion in Sect. \ref{sec:intro}, during the core collapse we may also use the fit to determine the core radius $\rc \equiv r_\I$.
Requiring function \eqref{eq:broken} to be continuous, the proportionality constants $b_{\II-\IV}$ must satisfy the conditions
\begin{equation}
  b_\II = b r_\I^{-a_\I + a_\II} \,,
\end{equation}
\vspace{-20pt}
\begin{equation}
  b_\III = b r_\I^{-a_\I + a_\II} r_\II^{-a_\II + a_\III}
,\end{equation}
and
\begin{equation}
  b_\IV = b r_\I^{-a_\I + a_\II} r_\II^{-a_\II + a_\III} r_\III^{-a_\III + a_\IV} \,,
\end{equation}
where we defined $b \equiv b_\I$ as it is now the only proportionality constant to fit.
We again applied the \citetalias{gen_evol} algorithm to minimise the sum of the squared absolute deviations in the radial density
\begin{equation}
  s^2_\rho = \sum_{i=0}^{k-1}{\left[ \rho_i - \rho(r_i) \right]^2}
\end{equation}
using parameters $a_{\I-\IV}$, $r_{\I-\III}$ and $b$.

\subsubsection{Limits}

The radial density profile \eqref{eq:broken} is expected to be a strictly decreasing function above the core radius. In addition, the slope $a_\III$ is supposed to have a value close to $\alpha$\,. Therefore, we defined common boundaries of $a_{\II-\IV} \in [0, 6)$. According to \citet{lb_e}, the radial density profile is flat in the core ($a_\I \approx 0$), but working with discrete data, binning has a great impact on the calculation of the Lagrangian shells. Therefore, we allowed for the density also having a positive slope in the core, giving it a generous range $a_\I \in (-6,6)$.
We took the inner and outermost Lagrangian radii for the limits of $r_{\I-\III}$, so $r_{\I-\III} \in (r_1, r_k)$. We also used an implicit condition that $r_\I < r_\II < r_\III$. A typical value of the proportionality constant was for example $b \approx 1.6$ in model \emod\ and $b \approx 0.8$ in model \mmod. Hence, we set the range of $b$ in all four models reasonably wide (e.g.\ from $-5$ to $5$) not to restrict the fits.

\section{Results}

The method we described above was applied to all of our models (\emod, \emodi, \mmod, \mmodi,\ and \mmodii). First, we discuss the noise reduction. Then, for each model individually, we focus on finding the minima, and fitting the time of core collapse and radial density profile. Finally, we compare our results with another method for estimating the time of core collapse\ via the formation of hard binary stars.

\subsection{Data preparation}

As we mentioned above in Sect. \ref{sec:method_cc}, oscillations of the unstable collapsing core make it impossible to seek systematic evolution of the Lagrangian radii using the raw data. We reduced those fluctuations using the \citetalias{savitzky_golay} algorithm with a second order polynomial, and the window width of the order of ten crossing times for each model. Results of this procedure are shown in the upper panels of Figs.~\ref{fig:lagr_s10k} to~\ref{fig:lagr_m100k}, where we plot the evolution of one arbitrarily chosen realisation of models \emod, \emodi, \mmod,\ and \mmodi. The effect of this data filtering is most visible in the inner Lagrangian radii of both plotted multi-mass models, where the noise is very high due to a small number of massive stars dominating the cluster's central region.

Alternatively, averaging the Lagrangian radii over many realisations (bottom four panels of Figs.~\ref{fig:lagr_s10k} to~\ref{fig:lagr_m100k}) also substantially reduces random fluctuations. For the multi-mass model, it renders the time of core collapse clearly visible as the global minimum of the inner Lagrangian radii. This may not be so obvious in a single realisation, where the global minima of the inner Lagrangian radii (either smoothed or not) may occur at later times (see the case presented in Figs.~\ref{fig:lagr_m20k}~\&~\ref{fig:lagr_m100k}).
However, the advantage of the \citetalias{savitzky_golay} filtering is that it not only allows us to identify $\tcc$ in individual realisations, based on the position of the local minima of the inner Lagrangian radii, but it also reveals long-term oscillations of the core region, which could be related to gravothermal oscillations in more populous models \citep{makino}.

\subsection{Equal-mass models}

In the case of model \emod, we have integrated each realisation twice, first for an overview of the global evolution, as shown in Fig.~\ref{fig:lagr_s10k}. When the contraction of the inner cluster region started, we reintegrated the models once more with a more frequent output until the contraction ended in order to have high resolution data for the following analyses. Filtering these outputs provided us with a clear single global minimum of each Lagrangian radius. The set \eqref{eq:lr_minima} was then constructed from these minima of the Lagrangian radii of mass fractions $m_{i, i \geq 1} \in [0.001,0.03]$ with a step of $0.001:$  those points are plotted for an arbitrarily chosen realisation in the top left panel of Fig.~\ref{fig:lagr_s}.
In this particular realisation, the fitting procedure described in Sect. \ref{sec:method_cc} gives $\tcc \approx 2317$ and $\alpha \approx 2.215$. After the evaluation of all one hundred realisations, we found that the dispersion of the fitted $\tcc$ indicates that the core collapse happened at a comparable time $2297 \pm 52$ (see Table~\ref{tab:Tcc_alpha}). This result is consistent with \citet{fujii_pz} who argued that the time of core collapse should be characteristic for any given model, depending only on the ratio between the mass of the most massive star and the mean stellar mass. Our fit of the power-law index $\alpha = 2.33 \pm 0.02$ (see Table~\ref{tab:Tcc_alpha}) also agrees with theoretical expectations \citep{lb_e,cohn,takahashi}.

At the time of core collapse, we constructed the radial density profile using the method described in Sect. \ref{sec:method_rho}. The density profile (see the bottom left panel of Fig.~\ref{fig:lagr_s} for an example) is in accord with our expectations: it is almost flat in the core, steeper around the core (slightly shallower at larger radii), and the steepest in the halo (see Table~\ref{tab:alpha} for the slopes and Table~\ref{tab:logr} for the radii where the density profile breaks). The fitted slope $a_\III \approx 2.321$ from \eqref{eq:broken} agrees with previous $N$-body simulations \citep[e.g.][]{core_collapse}. The value of $a_\III$ is nearly identical to the power-law index $\alpha \approx 2.33$ obtained by fitting the Lagrangian radii. The near equality of $\alpha$ and $a_\III$ indicates that the equal-mass model really evolved self-similarly before the core collapse. The outermost logarithmic density gradient, $a_\IV \approx 3.4$, is also in a good agreement with the prediction of \citet{spitzer_hart}.

After the first major contraction of the core region in model \emod,\ the inner Lagrangian radii show at least one post-collapse oscillation before the end of integration \citep[a similar behaviour was described also e.g. by][]{makino}. In such a small cluster, which barely exceeds the number of particles needed to form an unstable core \citep[cf.][]{goodman}, post collapse oscillations are not very deep and thus hard to detect. With a higher number of stars in a model, the core density increases and the collapse will be deeper \citep[e.g.][]{hut96,fujii_pz}. Due to these reasons, we have integrated an additional model (\emodi) of a more massive equal-mass star cluster, where those gravothermal oscillations are more easy to quantify. In order to determine the time of core collapse and the time of the subsequent contractions, we used the first and second time derivatives of each Lagrangian radius (calculated with finite differences) to identify the points of local minima. To eliminate false detection, we took a moving window with the same width as in the smoothing (of the order of ten crossing times) and kept only the deepest minimum in that window. Then we treated each of these sequences separately.
For the region around the core, we used mass fractions from $0.001$ to $0.020$ with a step of $0.001$ (see the first sequence in the top right panel of Fig.~\ref{fig:lagr_s}; the plotted realisation is one of the few that had two post-collapse oscillations in the integration window). At each minimum, we constructed the radial density profile of the cluster.

The first major contraction in \emodi\ occurred at time $\tcc \approx 9350$ (see Fig.~\ref{fig:lagr_s50k}); the subsequent contractions appear at a slightly different time and their depth varies in individual realisations. The homologous index reads $\alpha \approx 2.24$ at the core collapse in all realisations. This value is in agreement with \citet{lb_e} and comparable with the slope $a_\III$ (see Table~\ref{tab:alpha}). The radial density profile is in good agreement with model \emod\ and the predicted slopes \citep[cf.][]{lb_e,spitzer_hart}.
We also fitted the time of the first post-collapse oscillation in \emodi, which was present in all realisations of this model (e.g.\ see the second sequence in the top right panel of Fig.~\ref{fig:lagr_s}), and its corresponding radius density profile (the green curve in the bottom right panel of Fig.~\ref{fig:lagr_s}). The plots and the results printed in Tables~\ref{tab:Tcc_alpha} and~\ref{tab:alpha} show that in this case $\alpha \approx 2.34$ and $a_\III \approx 2.37$. Thus, we claim that the first post-collapse oscillation is homologous, as well as the core collapse.

\subsection{Multi-mass models}

The inner Lagrangian radii of both \mmod\ and \mmodi\ models show multiple low-frequency waves that represent several contractions of the core region in the time span of tens of crossing times. Those contractions are unevenly distributed in time in the individual realisations, that is\ they are not visible in the averaged radii (cf. upper and lower panels in Figs.~\ref{fig:lagr_m20k}~\&~\ref{fig:lagr_m100k}). Only the first contraction, which corresponds to the core collapse, seems to happen at the same time in all realisations ($\tcc \approx 53$ in \mmod\ and $\tcc \approx 116$ in \mmodi). Because the depths of these contractions vary, we used derivatives of the Lagrangian radii to determine the sequences of minima, as in the case of \emodi.
In 80\,\% of realisations we found two and in 45\,\% of realisations three clear consecutive minima. Due to a greater effect of binning, in this model we used larger mass fractions $m_{i, i \geq 1} \in [0.005,0.03]$ with a step of $0.005$ (see the top panels of Fig.~\ref{fig:lagr_m}) when constructing each set \eqref{eq:lr_minima}.

We fitted each of these sequences separately (see an example of this procedure in the right panels of Fig.~\ref{fig:lagr_m}; the corresponding mean values of $\tcc$ and $\alpha$ are listed in Table~\ref{tab:Tcc_alpha}). Although the points of minima are not as ordered as in the equal-mass models, the first contraction is well defined across all realisations and its deviation is rather small. The mean times of the second and third minima have higher uncertainties and their $1\sigma$ intervals overlap. This also proves that those minima are unevenly distributed in time across all realisations. The derived values of $\alpha$, about $1.5$, are significantly different from the theoretical predictions ($\alpha \approx 2.2$) but within their uncertainties, they are the same in all three minima.

For each estimated time of core collapse, we constructed the radial density profile (see Table~\ref{tab:alpha}). Qualitatively, it follows our expectations: almost zero in the centre, then a steep slope followed by a shallower one, and the steepest in the halo. However, the values of all slopes are different from those in the equal-mass models. Greater uncertainties that we see in the fits are inevitable due to a small number of particles that produce the Lagrangian radii of small mass fractions; each radius is highly influenced by any massive star that passes through the central region. Further, we note that the halo has a much steeper profile than predicted analytically. Nevertheless, the value of $a_\III \in (1.6, 1.7)$ (see Table~\ref{tab:alpha}) is similar in all minima and compatible with $\alpha$ within its uncertainties. This result indicates that the multi-mass clusters could evolve self-similarly too, albeit with different parameters than equal-mass or analytic models.

The detected post-collapse contractions of the cluster core most probably do not represent gravothermal oscillations because they are too separated in time and the systems are too small to form an unstable core \citep[e.g.][]{breen_heggie_two,breen_heggie_multi}. We rather refer to them as homologous collapses; they look almost the same and have comparable (and perhaps self-similar) properties. This leads us to conclude that in a more complicated system (e.g.\ a real star cluster, which we are well aware our models are not) simply analysing its post-collapse dynamical structure may not be enough to distinguish which collapse it has already sustained.

The depth of core collapse and its subsequent oscillations depends on the number of massive stars in the model and the ratio between the mass of the most massive star and the total mass of the cluster \citep{breen_heggie_two,breen_heggie_multi}. In both \mmod\ and \mmodi,\ this ratio was of the order of $10^{-3}$. In order to approach the depth of core collapse of our equal-mass models (where this ratio is $10^{-4}$ or $10^{-5}$), we made an additional multi-mass model (\mmodii) with the same initial mass function slope as models \mmod\ and \mmodi, but with a slightly modified range of masses to acquire a ratio of $10^{-4}$ (see Table~\ref{tab:models}). Even in this system, the collapse was not as deep as we expected. Fluctuations of the inner Lagrangian radii made our method for finding self-similar core collapse inefficient as can be seen for instance in the top panel of Fig.~\ref{fig:lagr_n} where our method detected two equivalent collapses of which only one is a potential core collapse. Based on the radial profile constructed at the times of minima, we got the power-law slope $a_\III \approx 1.9$ but the index $\alpha$ was different in most cases, ranging from $1.0$ to almost $3.2$, and in some cases, the sequence of minima was increasing instead of decreasing.

Self-similar evolution is a feature of collisionless systems. Although we found traces of it in some collisional systems, in others where for example close encounters are more dominant, we did not. Based on our results, we are unable to make a general statement on whether multi-mass models evolve self-similarly prior to the core collapse or not. In order to evaluate self-similar evolution in star clusters with a mass function, larger models with higher $N$ or a modified approach would be needed.

\subsection{Binary star formation}

The binding energy of a binary star (composed of stars with masses $m_a$ and $m_b$, with a semi-major axis $d$) is
\begin{equation}
  \Ebin \equiv \frac{m_a m_b}{2d} \,.
\end{equation}
We express the binding strength of a binary star in terms of $kT$, where $k$ is the Boltzmann constant and $T$ is the kinetic temperature, which relates to $N$-body variables (in Hénon units) by
\begin{equation}
  1\,kT = \frac{1}{6N} \,.
\end{equation}
Once a binary star becomes sufficiently tightly bound\ $(\Ebin$ exceeds several $kT$) it has a very low probability of being destroyed due to few-body interactions \citep{heggie,bin_stable}. To acquire that amount of energy, the pair must live in a dense environment, such as a collapsing cluster core \citep{tanikawa}.

It has been suggested \citep[e.g.][]{fujii_pz} that the time, $\tbin$, of the first appearance of a hard binary star with a critical energy
\begin{equation}
  \label{eq:fujii_pz}
  \Elim \equiv 10\, \frac{m_\mathrm{max}}{\langle m \rangle} \, kT \,,
\end{equation}
where $m_\mathrm{max}$ and $\langle m \rangle$ are the maximum and mean mass, respectively, identifies the core collapse. In the following, we test this hypothesis using our independent method for finding the time of core collapse.

For our model \emod, where $m_\mathrm{max} = \langle m \rangle$, Eq. \eqref{eq:fujii_pz} gives $\Elim = 10\,kT$. For that value of energy, we found a very good correlation between $\tbin$ and $\tcc$ (coefficient $0.884$, see Table~\ref{tab:corr} and left panel of Fig.~\ref{fig:Tcc_Tbin_s10k}). A slightly better correlation coefficient of $0.909$ was found for $\Elim = 100\,kT$, which indicates that the process of energy transfer into binaries is very quick, as was also pointed out by \citet{tanikawa}. During the process of core collapse (i.e.\ in the range of a few crossing times around $\tcc$) a typical binary star acquires energy $\Delta\Ebin$ of a few hundreds or even a thousand $kT$ (see the histogram in Fig.~\ref{fig:E_smod}), which is at least one order of magnitude greater than the binding energy which is supposed to identify the core collapse.

Equation \eqref{eq:fujii_pz} gives $\Elim \approx 750\,kT$ for the \mmod\ model. Following the same method as in \emod, we calculated the correlation coefficients for various binding energies of binary stars in the first collapse. Our choice of limiting binding energy comes from the histogram in Fig.~\ref{fig:E_mmod}, which shows the increase of binding energy of the hardest binary during each collapse. We found equivalent correlations for a range of energies between $750\,kT$ and $1250\,kT$ (compare the results in Table~\ref{tab:corr} and Fig.~\ref{fig:Tcc_Tbin_m20k}), which also supports the idea brought by \citet{tanikawa}.

In the case of model \mmodii, the limiting energy for a hard binary is $\Elim \approx 100\,kT$. In this model, we have detected several binary stars with $\Ebin > \Elim$ existing at $t \approx \tcc$ (see the lower panel of Fig.~\ref{fig:lagr_n}). The first major contraction in the plotted realisation was driven by several binaries and the second seems to form a $13\,000\,kT$ binary star (which is well above $\Elim$). Due to a high uncertainty in determining the time of core collapse in this model, we were unable to calculate the correlation with the time of formation of the first hard binary star.

Let us also point out that rapid hardening of binaries, or formation of new ones, is not unique to the core collapse. We found several occurrences of such events even though we did not find any sign of collapse in the star cluster (e.g.\ compare the process of binary evolution in models \mmod\ and \mmodi\ in the bottom panels of Fig.~\ref{fig:lagr_m}).
On the left hand side, we see the evolution of two hard binaries (plotted in red and blue) in model \mmod. The red binary star formed during the first collapse of the core and hardened continuously. In the subsequent homologous collapse, a short-lived binary star (green) emerged and the existing one (red) acquired almost twice its former binding energy. A detailed analysis showed that the disappearance of the green binary was due to a close interaction with the red one at that time. During the third collapse, another hard binary (blue) started to form and harden. The event of large energy change of a binary that occurred out of sync with any collapse is clearly visible on the red evolution track.

In the bottom right panel of Fig.~\ref{fig:lagr_m}, we show a similar process in one realisation of model \mmodi. Although the formation of the first binary (red) began earlier than the core collapse determined from the fitting, the first major increase of its binding energy is linked to the first contraction of the core. A detailed look at the output revealed that a new binary (green), which formed at $t \approx 200$, interacted at $t \approx 250$ with the red binary star. This caused an immediate hardening of the final binary, which follows in red, and interchange of the components. During the second homologous collapse in the plotted realisation, the red binary hardened and it was expelled from the cluster, and a new binary (blue) started to form. It gradually hardened and is likely to be related to the third homologous collapse. Also in this realisation, we see cases of formation of binaries (the green one) or their rapid hardening (the blue binary after time 450) that are out of sync with the collapses of the core.

The correlation between the time of core collapse and the time of the formation of the first binary star in the more massive equal- and multi-mass models (i.e.\ \emodi\ and \mmodi) is based on the data from only ten realisations. Hence, characterising a typical energy of a binary star that was promoted by the core collapse is influenced too much by the statistical noise. We do not draw any conclusions from these particular results. Nevertheless, we may observe a similar trend as in the smaller models, that is\ that the core collapse is linked with the formation of very hard binary stars. In the case of \emodi\ (see Fig.~\ref{fig:Tcc_Tbin_s50k}), the best correlation is achieved for $\Elim$ between $100\,kT$ and $1000\,kT$ with a coefficient above $0.98$.
In the case of model \mmodi, the correlation coefficients vary from $0.47$ to $0.78$ for $\Elim$ above $1500$ or $10^4\,kT$, respectively (compare also the plots in Fig.~\ref{fig:Tcc_Tbin_m100k}). These energies are again well above the estimate from Eq. \eqref{eq:fujii_pz}.

We did not make an attempt to evaluate the correlation of formation (or hardening) of binaries with the second or third homologous collapse as it is virtually impossible to distinguish which collapse is responsible for creating a binary (or vice versa) after the system has already collapsed once.

\section{Conclusions}

We investigated the properties of core collapse in numerical $N$-body models of self-gravitating star clusters. For that purpose, we developed a novel method
for the identification of the time of core collapse. The method is based on an assumption proposed for analytic models by \citet{lb_e} that the evolution of the cluster is self-similar.

In the case of equal-mass models (\emod\ and \emodi), we found a very good agreement with theoretical expectations. Minima of the Lagrangian radii for small mass fractions are aligned according to a power-law relation $\rlmin \propto \big( \tcc - \tmin \big)^{\frac{2}{6 - \alpha}}$ with the power-law index close to $\alpha \approx 2.3$. At the time of core collapse, the cluster's radial density profile 
in the intermediate region between the core and the half-mass radius is well approximated by a power law $\rho \propto r^{-a_\III}$, with $a_\III \approx 2.3$. The fact that $a_\III \approx \alpha$ indicates that the cluster's evolution matches the self-similar solution of \citet{lb_e}.
The density profile in the halo is best fitted by a power law with index $a_\IV \approx 3.4,$ which is close to the prediction formulated by \citet{spitzer_hart} for the evolution of halos of $N$-body models.

Further, we analysed $N$-body models of star clusters with a \citet{salpeter} mass function (\mmod\ and \mmodi). Using our method, we identified the times of core collapse and determined the radial density profile of the clusters at that moment. We found that the cluster's evolution and density profile are qualitatively similar to the previous case, although the power-law index $\alpha$ has a significantly different value. Specifically, the best-fit value of $\alpha$ for temporal evolution of the inner Lagrangian radii is $1.5$, which is nearly identical to the power-law index of the radial density profile beyond the cluster core, $a_\III \approx 1.6$. Thus, we conclude that these models show traces of self-similar evolution.

We also studied the evolution of a multi-mass model (\mmodii) with the same slope of the mass function as \mmod\ and \mmodi\ but a higher ratio between the total mass and the most massive star. In terms of self-similar evolution, we expected this model to be a ``bridge'' between the equal- and multi-mass models that we have already discussed. However, there were big differences in the radial profiles across the realisations, caused by random oscillations of the core region. In most realisations, we were unable to successfully fit the minima of the Lagrangian radii and clearly determine the time of core collapse and its homologous properties.

Our results show that analytic predictions on the self-similar evolution are valid in the limit of equal-mass $N$-body systems but cannot be straightforwardly generalised for multi-mass (i.e.\ more realistic) star clusters. A further study from both the analytic and numerical point of view is needed to conclude whether multi-mass systems with a general mass function do undergo self-similar core collapse evolution, perhaps with the homologous index dependent on the mass function properties. Any future studies of this topic would certainly benefit from analysing even more populous clusters.

In the case of \mmod\ and \mmodi\ as well as in \emodi, we found subsequent phases of coherent evolution of the inner Lagrangian radii even after the core collapse. Evolution toward all those minima have similar characteristics and homologous properties (i.e.\ depth of the core contraction, power-law indices $\alpha$ , and the radial density profiles). Therefore, we conclude that they are observationally indistinguishable from each other. The only prominent difference between the first and subsequent homologous collapses is that the time of the first one is well correlated among different realisations, which corresponds to the argument made by \citet{fujii_pz}. Our values of $\tcc$ are $52.9 \pm 8.1$ (\mmod) and $116 \pm 38$ (\mmodi), while for the second and third collapses, for example\ in \mmod, we have found $101 \pm 21$ and $120 \pm 17$, respectively (see also Table~\ref{tab:Tcc_alpha}). A large deviation of the times of subsequent collapses implies that they are smeared out in the plots of the Lagrangian radii averaged over all realisations of the particular models (see Figs.~\ref{fig:lagr_m20k}~\&~\ref{fig:lagr_m100k}). In the case of \mmod, we identified two such homologous collapses (including the first one) in 80\,\% and three in 45\,\% of the realisations within the integration time. All realisations of \mmodi\ passed at least three homologous collapses.

Finally, we studied the correlation of the time of core collapse, $\tcc$, determined by our method with the formation of dynamical binaries in the cluster. In the case of \emod, we found the best correlation of $\tcc$ with the time when the first binary acquired binding energy higher than $100\,kT$ (correlation coefficient of $0.909$), yet only a slightly smaller value ($\approx 0.88$) was obtained for the correlation with the first occurrence of a binary with $\Ebin > 10\,kT$. This indicates that (i) the flow of energy toward the binaries is indeed very fast during the core collapse and (ii) the formation of the first hard binary with relatively poorly constrained binding energy may be used to identify the core collapse. 
In the multi-mass model we have the best correlations for binding energies between $750\,kT$ and $1250\,kT$, where the correlation coefficient is in the range from $0.53$ to $0.58$. Analytic estimates for the binding energy of binaries formed during the core collapse derived by \citet{fujii_pz} give values of $10\,kT$ and $750\,kT$ for the \emod\ and \mmod\ models, respectively.

Detailed inspection of our models revealed that a large transfer of binding energy from the cluster to binaries occurs not only during the core collapse but also during the subsequent homologous collapses. On the other hand, tracking the binding energy of binaries in our models (\mmod\ in particular) revealed that episodes of large energy transfer are much more numerous than the homologous collapses. In other words, there are common events of formation of (or hardening of existing) binaries that cannot be identified with any homologous collapse. In some cases, these interactions led to a change of binding energy of the order of $10^4\,kT$ on a timescale shorter than one crossing time,\ exceeding by an order of magnitude the energy transfer rate related to the homologous collapses.

The formation of the first hard binary star (in a system without primordial binaries) is well correlated with the phase of core collapse in a star cluster. As there are no other hard binaries present in the system, it is a good indicator of this event. After the system has already collapsed once and has produced at least one hard binary star, neither the formation of a new hard
binary nor a large transfer of binding energy into existing ones can be considered as an indicator of the subsequent homologous contractions. From that perspective, it would be intriguing to examine homologous properties and binary evolution during the core collapse in systems containing a primordial binary population.

\begin{acknowledgements}
First and foremost we thank Douglas Heggie for introducing us to this topic and for his selfless guidance throughout the whole project.
We sincerely thank the University of Edinburgh for its hospitality while VP was a visiting research student there for two months in 2016.
Many thanks to Sverre Aarseth for his kind help with \texttt{nbody6} and to Steve Shore for his perpetual support and thoughtful comments while finalising the paper.

VP and this study was mainly supported by Charles University, grants GAUK-186216 and SVV-260441.
LŠ also acknowledges support from the Czech Science Foundation through the project of Excellence No.\ 14-37086G.

We greatly appreciate access to the computing and storage facilities owned by parties and projects contributing to the National Grid Infrastructure MetaCentrum, provided under the programme \emph{Projects of Large Research, Development, and Innovations Infrastructures} (CESNET LM2015042). Large calculations were also performed on \emph{Tiger} at the Astronomical Institute of Charles University.

We thank the anonymous referee for useful comments.
\end{acknowledgements}

% for the bibliography, at the end
\bibliographystyle{aa}
\bibliography{Fitting_self_similar_core_collapse_to_N_body_models}

\clearpage
\appendix
\onecolumn
\section{Tables \& Figures}

%% tables
\begin{table}[h]
\begin{minipage}[t]{.49\linewidth}
  \centering
  \caption{Correlation coefficient between the time of core collapse, $\tcc$, and the time of formation of the first binary with binding energy $\Ebin \geq \Elim$ (i.e.\ $\tbin$, in our models; see also Figs.~\ref{fig:Tcc_Tbin_s10k} \&~\ref{fig:Tcc_Tbin_m20k}). The numbers $\tcc$ and $\tbin$ are given with their $1\sigma$ deviations calculated from the corresponding number of realisations.}
  \label{tab:corr}
  \begin{tabular}{ccccc}
    \hline
    model & $\tcc$ & $\Elim\ [kT]$ & $\tbin$                & $\left(\tcc , \tbin\right)$ \\
    \hline
    
    \emod & $2297 \pm 52$ &   10   &  $2279 \pm 59$         & $0.884$ \\
                   &               &  100 &  $2286 \pm 57$  & $0.909$ \\
                   &&&&\\
    \mmod & $52.9 \pm 8.1$ &  750  &  $52.1 \pm 6.9$        & $0.532$ \\
                   &               & 1000 &  $53.8 \pm 6.8$ & $0.576$ \\
                   &               & 1250 &  $55.2 \pm 6.7$ & $0.579$ \\
                   &               & 2000 &  $58.9 \pm 8.8$ & $0.481$ \\
    \hline
  \end{tabular}
% \end{table}
\end{minipage}
\hfill %\hspace{2pc}
\begin{minipage}[t]{.49\linewidth}
% \begin{table}[h]
  \centering
  \caption{Fitted mean time of core collapse and the power-law index $\alpha$ of the core radius temporal evolution. In the case of multiple homologous collapses, \# indicates their sequence number.  All numbers are given with their $1\sigma$ deviations calculated from the corresponding number of realisations.}
  \label{tab:Tcc_alpha}
  \begin{tabular}{cccc}
    \hline
    model & \# & $\tcc$ & $\alpha$ \\
    \hline
    
    \emod
    &                 & $2297 \pm 52$  & $2.33 \pm 0.02$ \\
    &&&\\
    \emodi
    & $1^\mathrm{st}$ & $9347 \pm 150$ & $2.24 \pm 0.16$ \\
    & $2^\mathrm{nd}$ & $9575 \pm 118$ & $2.34 \pm 0.42$ \\
    &&&\\
    \mmod
    & $1^\mathrm{st}$ & $52.9 \pm 8.1$ & $1.55 \pm 0.14$ \\
    & $2^\mathrm{nd}$ & $101 \pm 21$   & $1.50 \pm 0.18$ \\
    & $3^\mathrm{rd}$ & $120 \pm 17$   & $1.50 \pm 0.18$ \\
    &&&\\
    \mmodi
    & $1^\mathrm{st}$ & $116 \pm 38$   & $1.55 \pm 0.16$ \\
    & $2^\mathrm{nd}$ & $235 \pm 34$   & $1.41 \pm 0.18$ \\
    & $3^\mathrm{rd}$ & $395 \pm 61$   & $1.40 \pm 0.23$ \\

    \hline
  \end{tabular}
\end{minipage}
\end{table}

\begin{table}[h]
        \centering
  \caption{Mean logarithmic density gradients, $a_{\I-\IV}$, of our models at a given collapse. The fitted radii where the power law breaks, $r_{\I-\III}$, are shown in Table~\ref{tab:logr}. All numbers are given with their $1\sigma$ deviations calculated from the corresponding number of realisations.}
  \label{tab:alpha}
  \begin{tabular}{cccccc}
    \hline
    model & \# & $a_\I$ & $a_\II$ & $a_\III$ & $a_\IV$ \\
    \hline
    
    \emod
    &                 & $0.60 \pm 0.22$ & $2.39 \pm 0.26$ & $2.32 \pm 0.07$ & $3.44 \pm 0.03$ \\
    &&&&&\\
    \emodi
    & $1^\mathrm{st}$ & $0.56 \pm 0.21$ & $2.41 \pm 0.03$ & $2.27 \pm 0.03$ & $3.36 \pm 0.01$ \\
    & $2^\mathrm{nd}$ & $0.61 \pm 0.15$ & $2.37 \pm 0.12$ & $2.19 \pm 0.02$ & $3.35 \pm 0.01$ \\
    &&&&&\\
    \mmod
    & $1^\mathrm{st}$ & $0.17 \pm 0.21$ & $2.09 \pm 0.60$ & $1.60 \pm 0.10$ & $3.96 \pm 0.08$ \\
    & $2^\mathrm{nd}$ & $0.12 \pm 0.24$ & $2.07 \pm 0.68$ & $1.70 \pm 0.09$ & $3.89 \pm 0.11$ \\
    & $3^\mathrm{rd}$ & $0.12 \pm 0.24$ & $2.22 \pm 0.64$ & $1.68 \pm 0.08$ & $3.80 \pm 0.10$ \\
    &&&&&\\
    \mmodi
    & $1^\mathrm{st}$ & $0.26 \pm 0.31$ & $1.76 \pm 0.24$ & $1.53 \pm 0.18$ & $3.97 \pm 0.07$ \\
    & $2^\mathrm{nd}$ & $0.64 \pm 0.25$ & $1.88 \pm 0.60$ & $1.64 \pm 0.10$ & $3.85 \pm 0.12$ \\
    & $3^\mathrm{rd}$ & $0.47 \pm 0.16$ & $1.95 \pm 0.21$ & $1.63 \pm 0.06$ & $3.61 \pm 0.10$ \\
    
    \hline
  \end{tabular}
\end{table}

\begin{table}[h]
        \centering
  \caption{Radii where the power law breaks, $r_{\I-\III}$, are shown as logarithms to be consistent with the figures. The mass contained in a sphere of the corresponding radius is $m(r_{\I}) \lesssim 0.01$, $m(r_{\II}) \approx 0.03,$ and $m(r_{\III}) \approx 0.40$ approximately in all models. All numbers are given with their $1\sigma$ deviations calculated from the corresponding number of realisations.}
  \label{tab:logr}
  \begin{tabular}{ccccc}
    \hline
    model & \# & $\log{r_\I}$ & $\log{r_\II}$ & $\log{r_\III}$ \\
    \hline
    
    \emod
    &                 & $-2.43 \pm 0.17$ & $-1.54 \pm 0.21$ & $-0.02 \pm 0.09$ \\
    &&&&\\
    \emodi
    & $1^\mathrm{st}$ & $-2.68 \pm 0.11$ & $-1.95 \pm 0.30$ & $-0.13 \pm 0.05$ \\
    & $2^\mathrm{nd}$ & $-2.59 \pm 0.06$ & $-1.84 \pm 0.14$ & $\ \ 0.16 \pm 0.03$ \\
    &&&&\\
    \mmod
    & $1^\mathrm{st}$ & $-1.26 \pm 0.09$ & $-0.97 \pm 0.09$ & $-0.15 \pm 0.05$ \\
    & $2^\mathrm{nd}$ & $-1.20 \pm 0.10$ & $-0.96 \pm 0.11$ & $-0.08 \pm 0.03$ \\
    & $3^\mathrm{rd}$ & $-1.18 \pm 0.09$ & $-0.94 \pm 0.13$ & $-0.07 \pm 0.03$ \\
    &&&&\\
    \mmodi
    & $1^\mathrm{st}$ & $-1.17 \pm 0.04$ & $-0.94 \pm 0.05$ & $-0.17 \pm 0.04$ \\
    & $2^\mathrm{nd}$ & $-1.20 \pm 0.09$ & $-0.99 \pm 0.11$ & $-0.10 \pm 0.03$ \\
    & $3^\mathrm{rd}$ & $-1.21 \pm 0.07$ & $-0.94 \pm 0.10$ & $-0.07 \pm 0.02$ \\

    \hline
  \end{tabular}
\end{table}

\clearpage
\twocolumn
%% figures
%% These 4 figures (8 plots) MUST stay on the same page
\begin{figure} %
        \centering  
        \includegraphics[width=.98\linewidth]{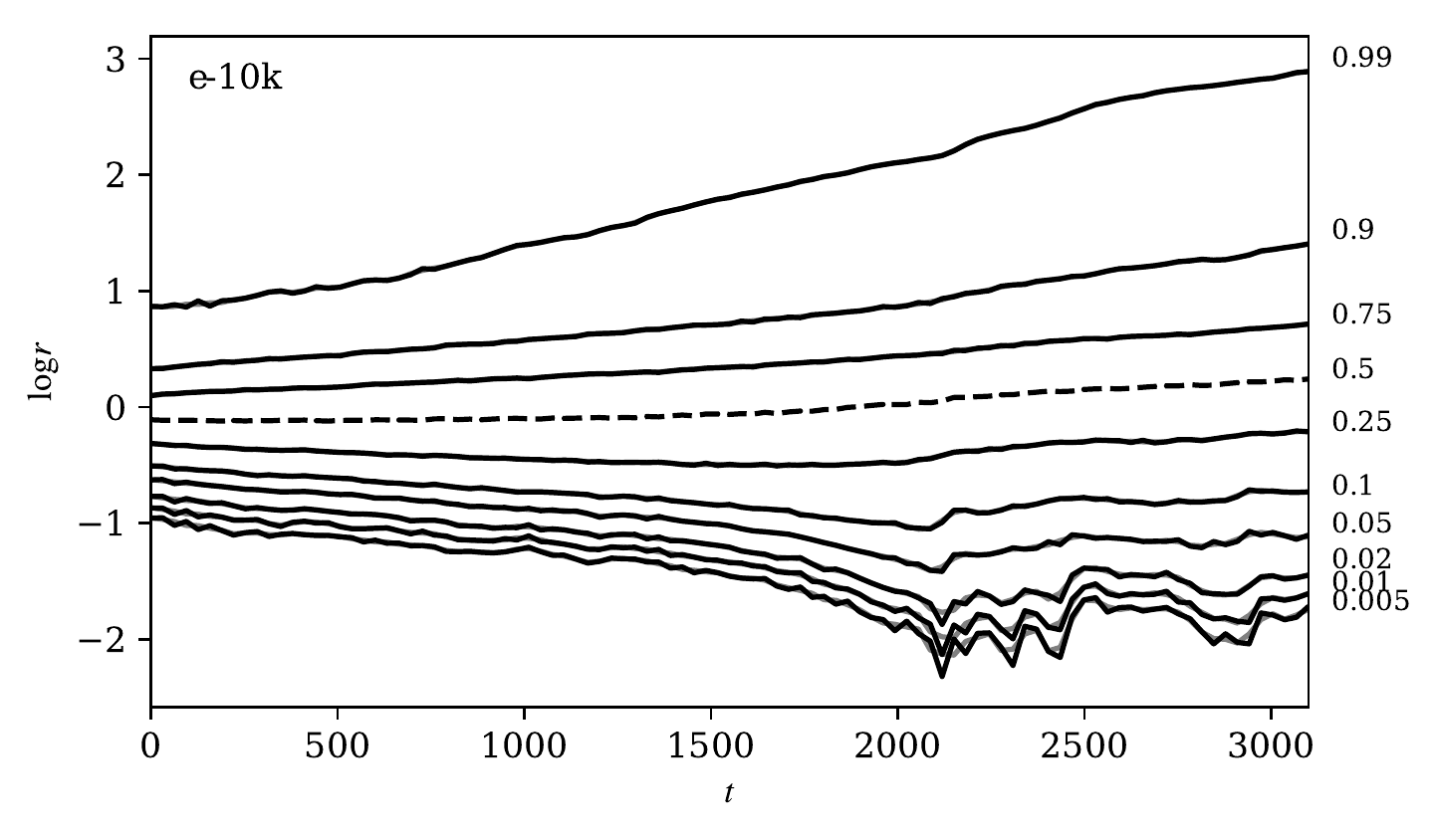}
        \includegraphics[width=.98\linewidth]{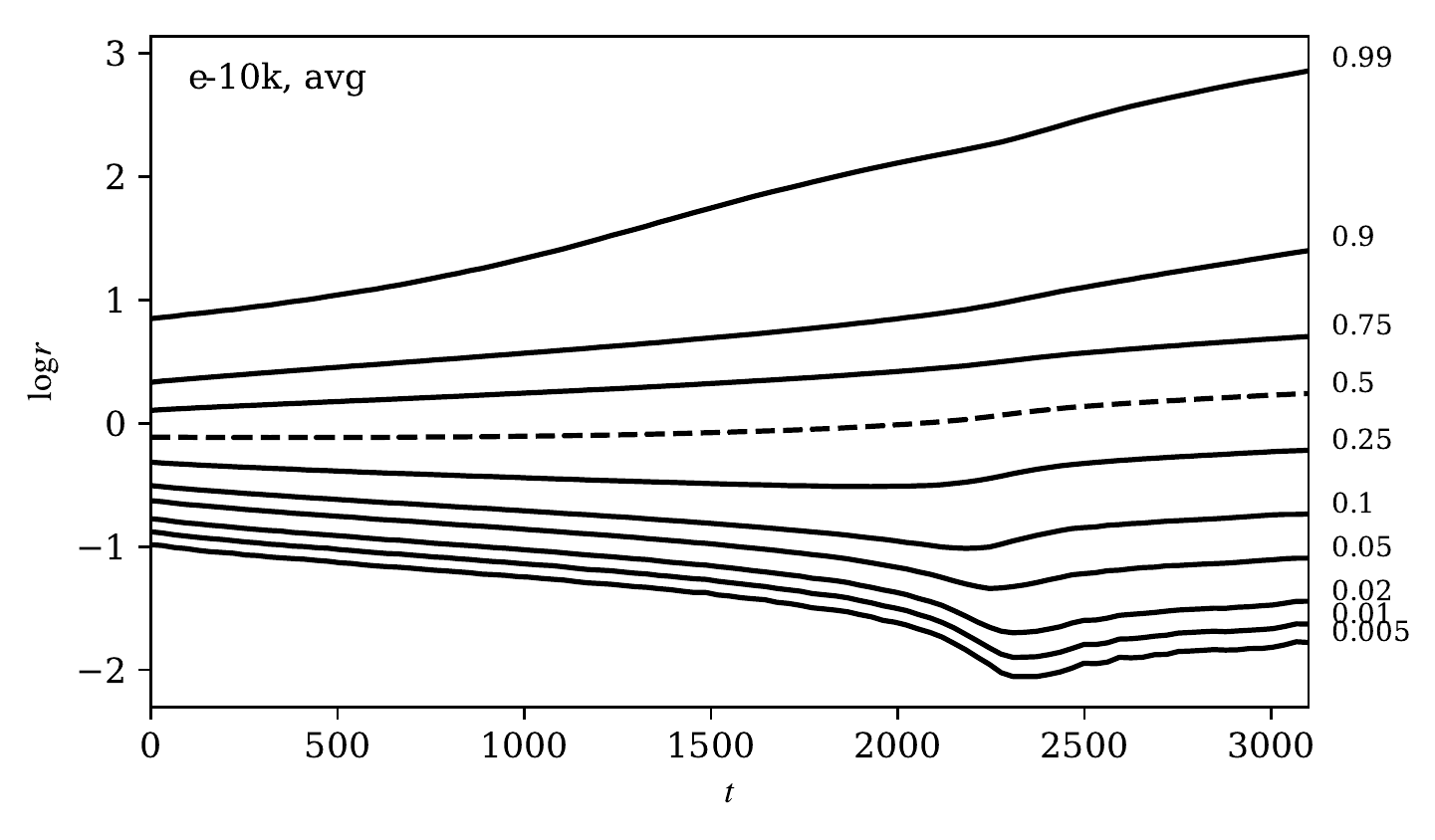}
        \caption{Lagrangian radii of model \emod. The mass fractions are on the right, the half-mass radius is plotted with a dashed line.
        \textbf{Top:} One arbitrary realisation. Smoothed curves are plotted in black, the original data are in grey.
        \textbf{Bottom:} Average over all realisations.}
        \label{fig:lagr_s10k}
\end{figure}

\begin{figure}
        \centering  
        \includegraphics[width=.98\linewidth]{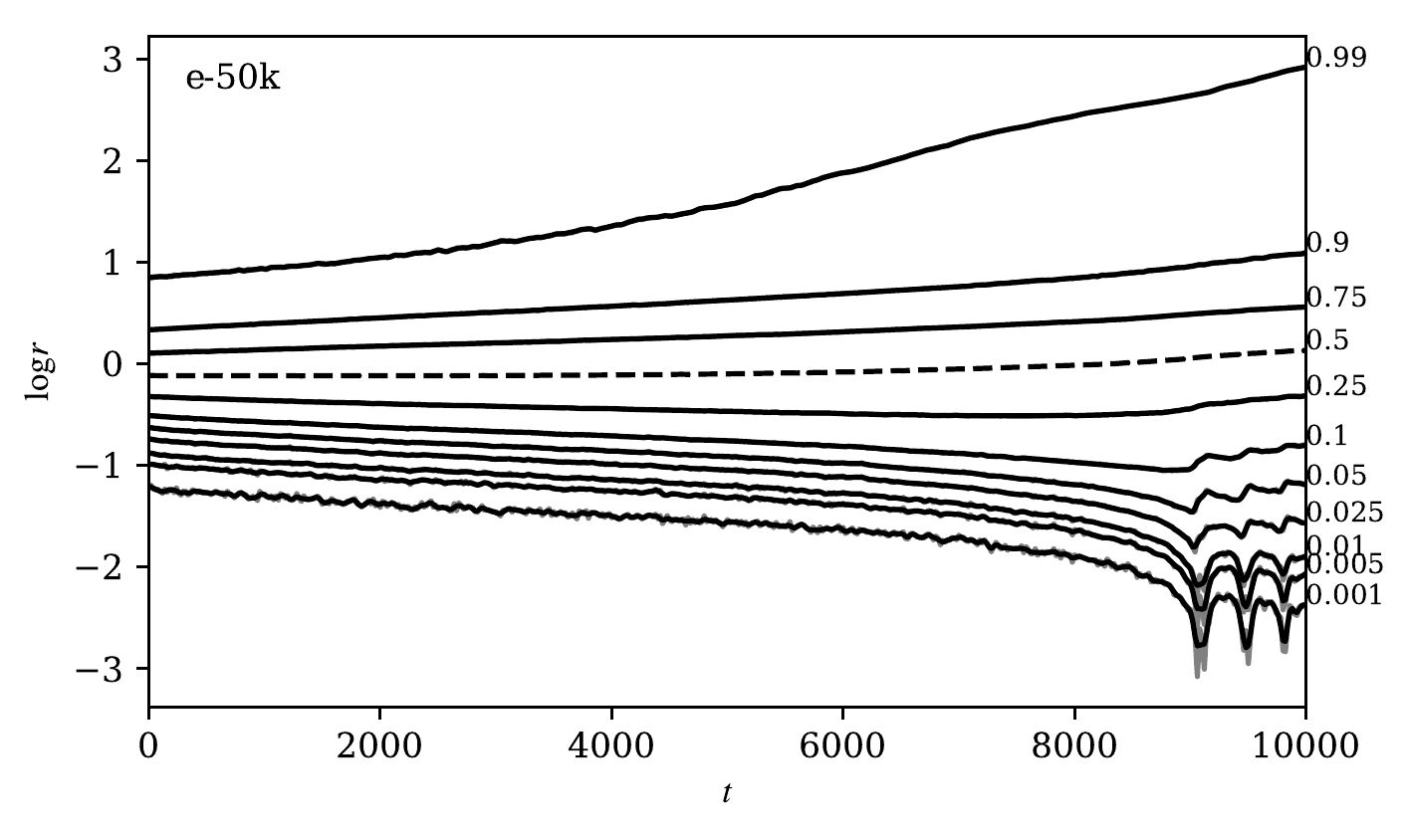}
        \includegraphics[width=.98\linewidth]{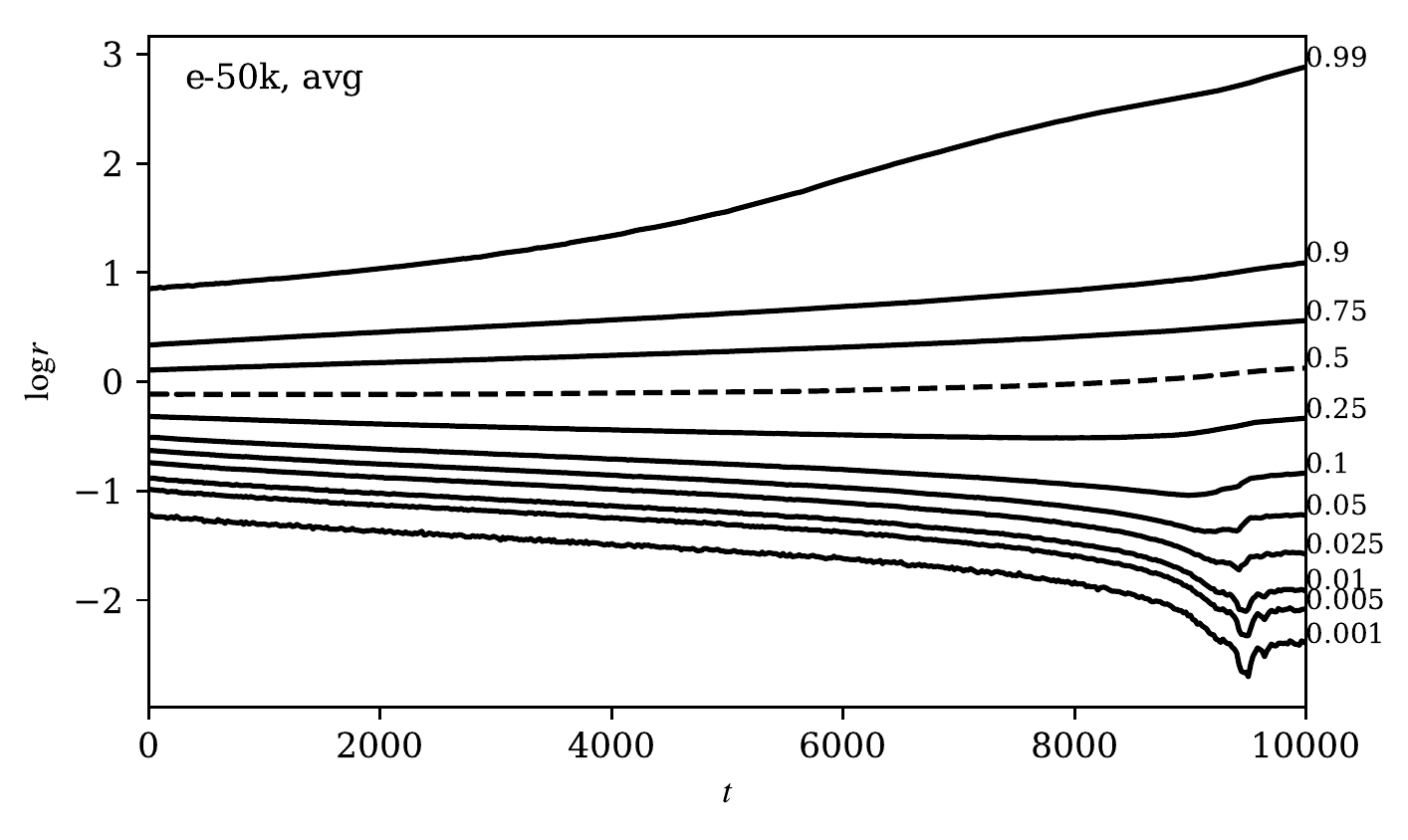}
        \caption{Same as in Fig.~\ref{fig:lagr_s10k} but for model \emodi.}
        \label{fig:lagr_s50k}
\end{figure}

\begin{figure}
        \centering  
        \includegraphics[width=.98\linewidth]{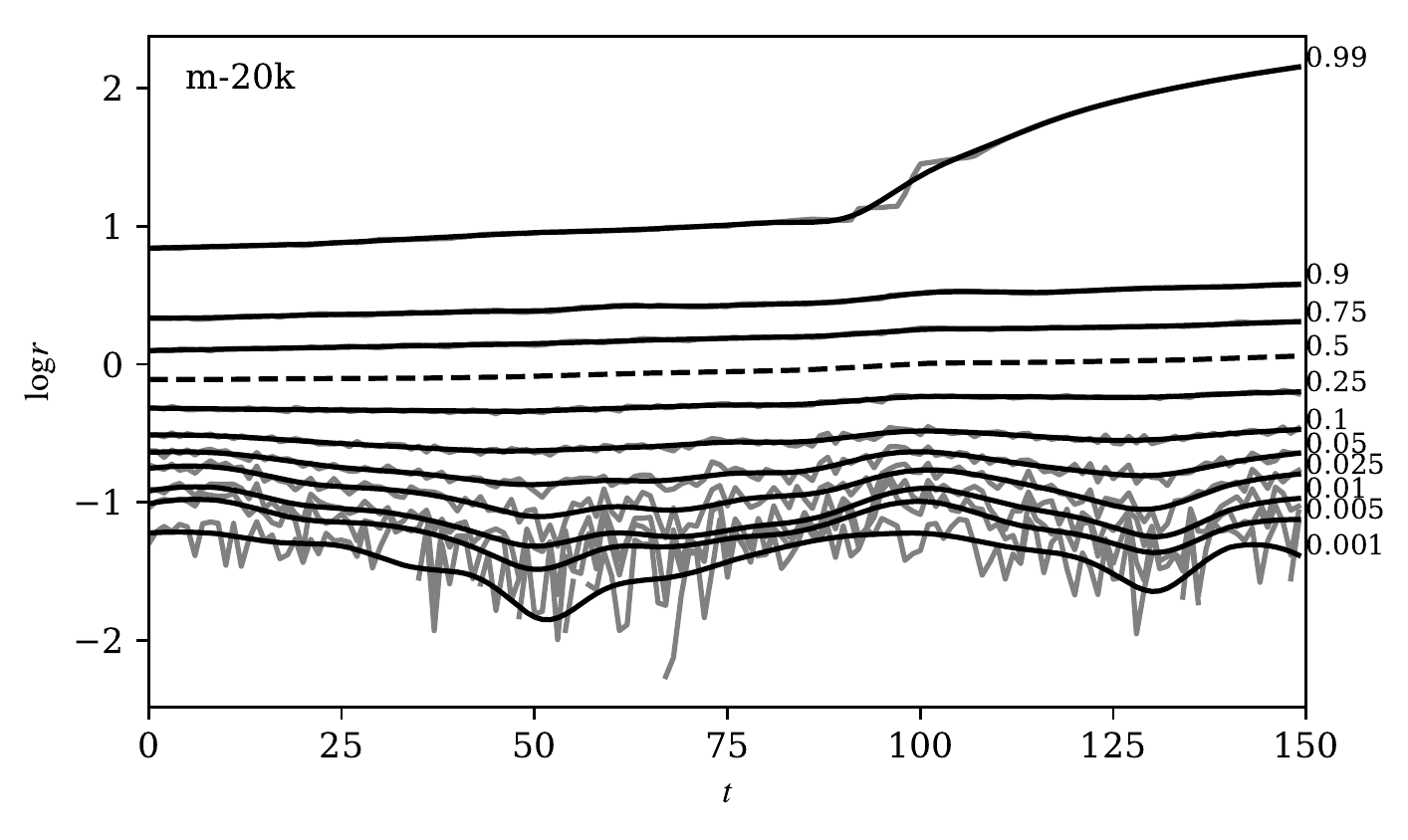}
        \includegraphics[width=.98\linewidth]{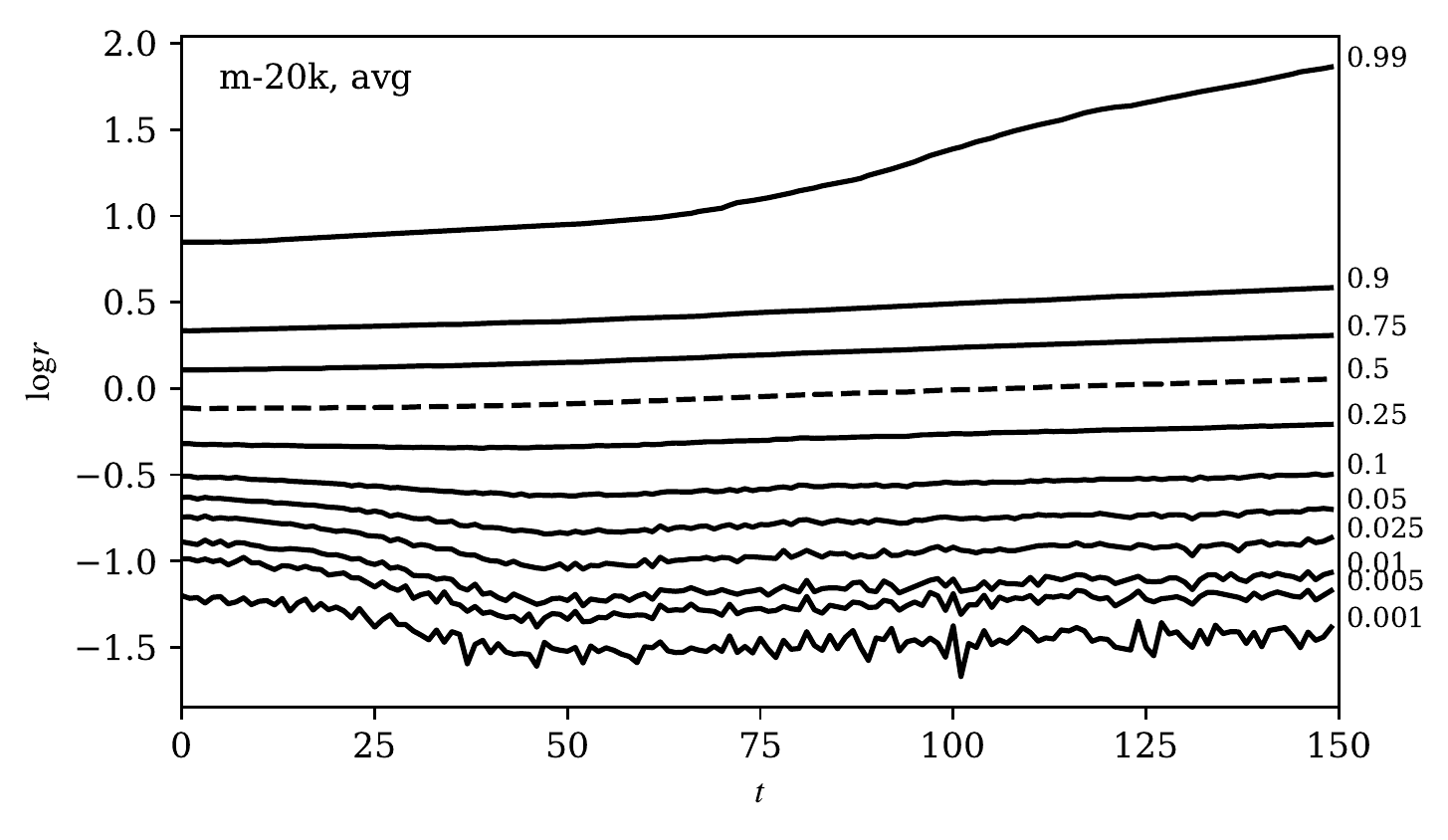}
        \caption{Same as in Fig.~\ref{fig:lagr_s10k} but for model \mmod.
        \vspace{28pt}} %% formating on a page
        \label{fig:lagr_m20k}
\end{figure}

\begin{figure}
        \centering  
        \includegraphics[width=.98\linewidth]{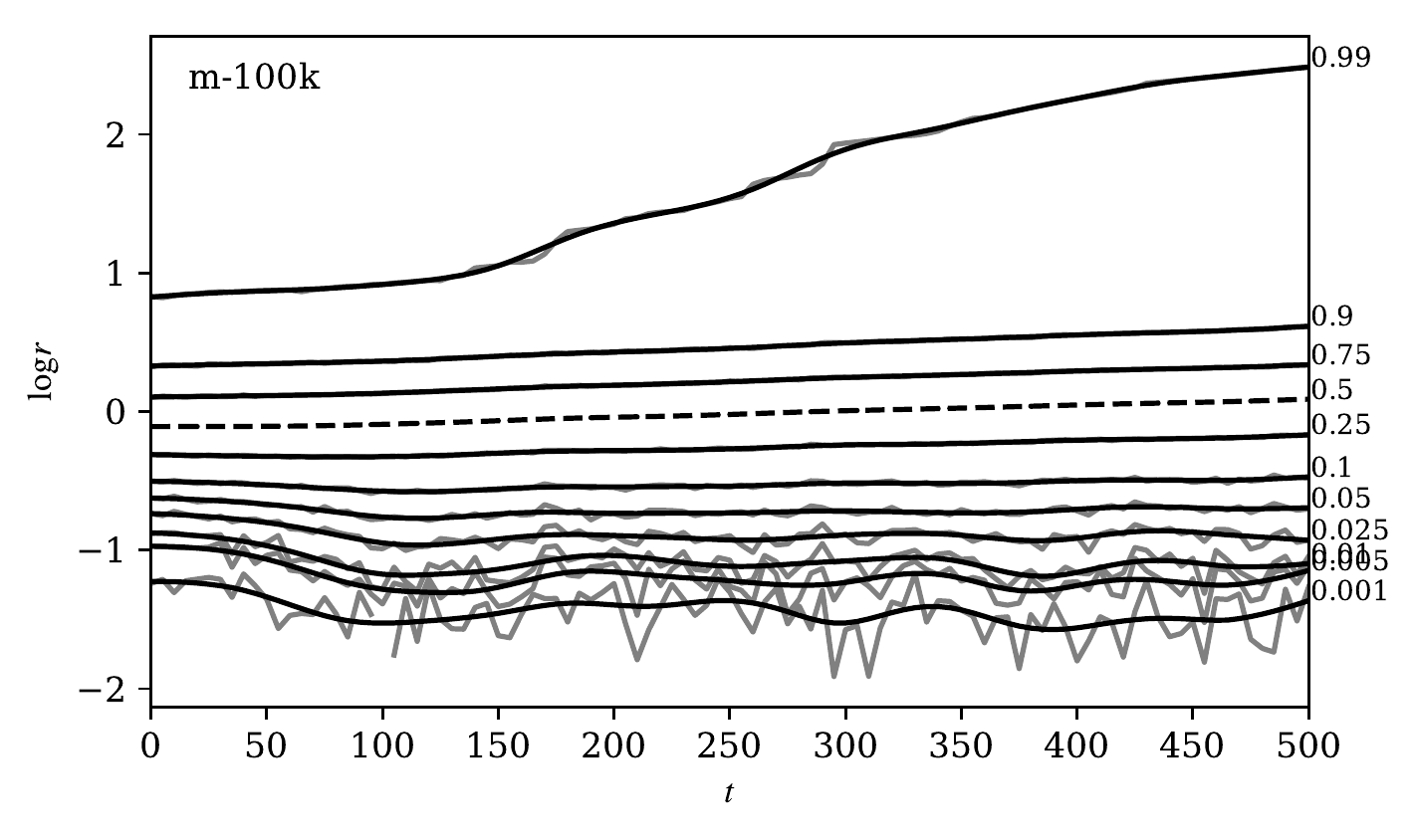}
        \includegraphics[width=.98\linewidth]{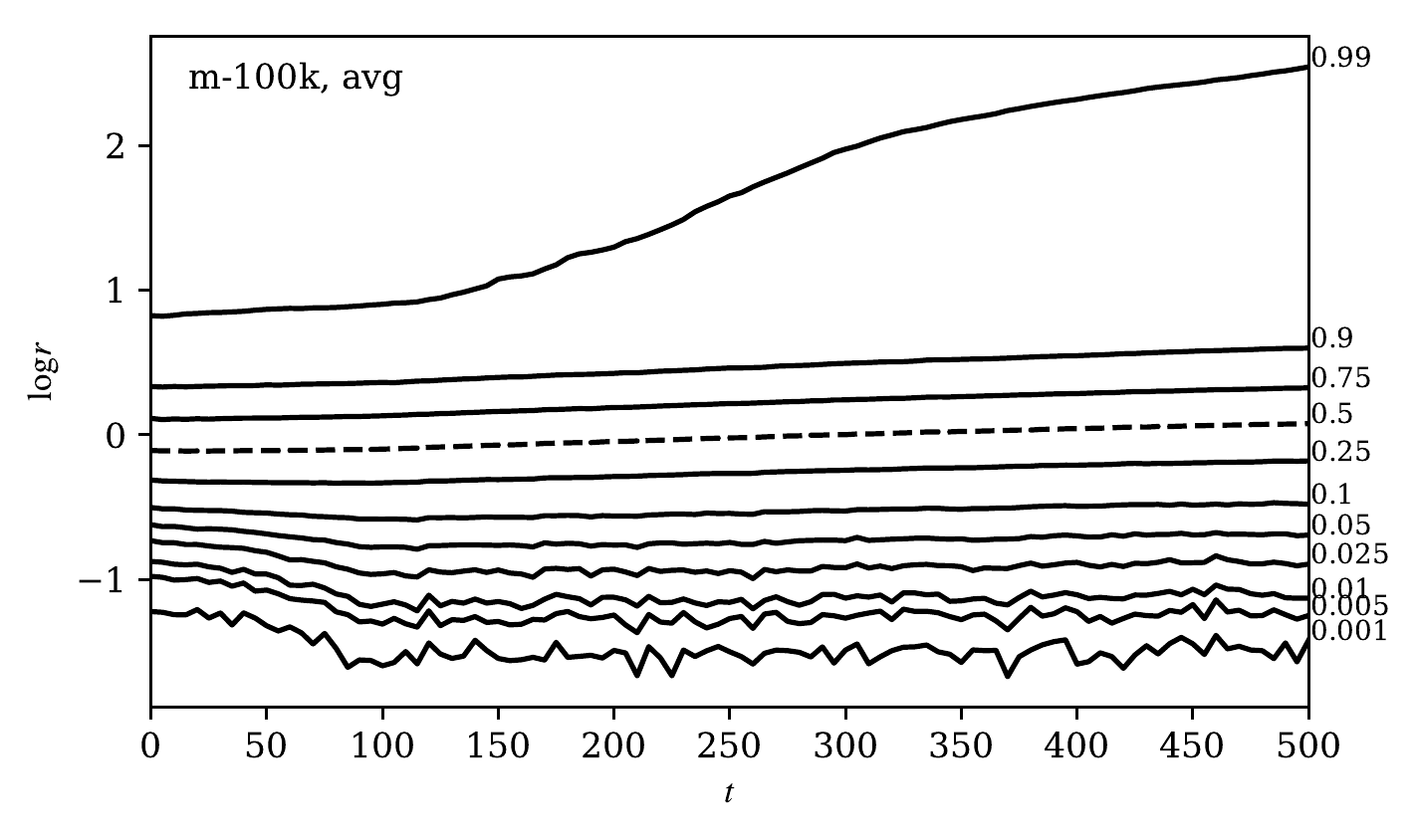}
        \caption{Same as in Fig.~\ref{fig:lagr_s10k} but for model \mmodi.}
        \label{fig:lagr_m100k}
\end{figure}

%%%
\begin{figure*}
  \centering
  \includegraphics[width=.49\linewidth]{{{ss_mod018-run002.p-eps-converted-to}}}\hfill
  \includegraphics[width=.485\linewidth]{{{ss_nbody011-run08.f-eps-converted-to}}}
  
  \caption{\textbf{Top:} Detail of the inner Lagrangian radii of one realisation of model \emod\ (\textbf{left}) and of model \emodi\ (\textbf{right}). Circles correspond to the minima of smoothed radii, that is,\ set \eqref{eq:lr_minima}, and the dashed line is a power-law fit \eqref{eq:fit_heaviside} through these data.\newline
  \textbf{Bottom:} Radial density profiles of this realisation at the given times in the range of radii plotted above. The dotted line demonstrates a fit by the triple-broken power-law function \eqref{eq:broken}. The grey line shows an initial state of the system for comparison.}
  \label{fig:lagr_s}
\end{figure*}

\begin{figure*}
  \centering
  \includegraphics[width=.49\linewidth]{{{ss_nbody008-run17.f-eps-converted-to}}}\hfill
  \includegraphics[width=.48\linewidth]{{{ss_nbody009-run04.f-eps-converted-to}}}
  
  \caption{Sequences of minima and binary star binding energy evolution in one realisation of model \mmod\ \textbf{(left)} and of model \mmodi\ \textbf{(right)}.\newline
  \textbf{Top:} Lagrangian radii. Circles correspond to the minima of the radii, and the dashed lines are the power-law fits \eqref{eq:fit_heaviside}.\newline
  \textbf{Bottom:} Evolution of the binding energies of the dynamically formed binary stars. Each colour corresponds to one binary star family line, in which the individual components may be exchanged due to interactions with other stars. Black dashed vertical lines indicate the times of the homologous collapses, highlighted areas around them are the time intervals used for the evaluation of $\Delta\Ebin$.}
  \label{fig:lagr_m}
\end{figure*}

\begin{figure}
  \centering
  \includegraphics[width=\linewidth]{{{ss_nbody013-run00.f-eps-converted-to}}}
  
  \caption{Same as in \ref{fig:lagr_m} but for one realisation of model \mmodii.}
  \label{fig:lagr_n}
\end{figure}
%%%

\begin{figure}
  \centering
  \includegraphics[width=\linewidth]{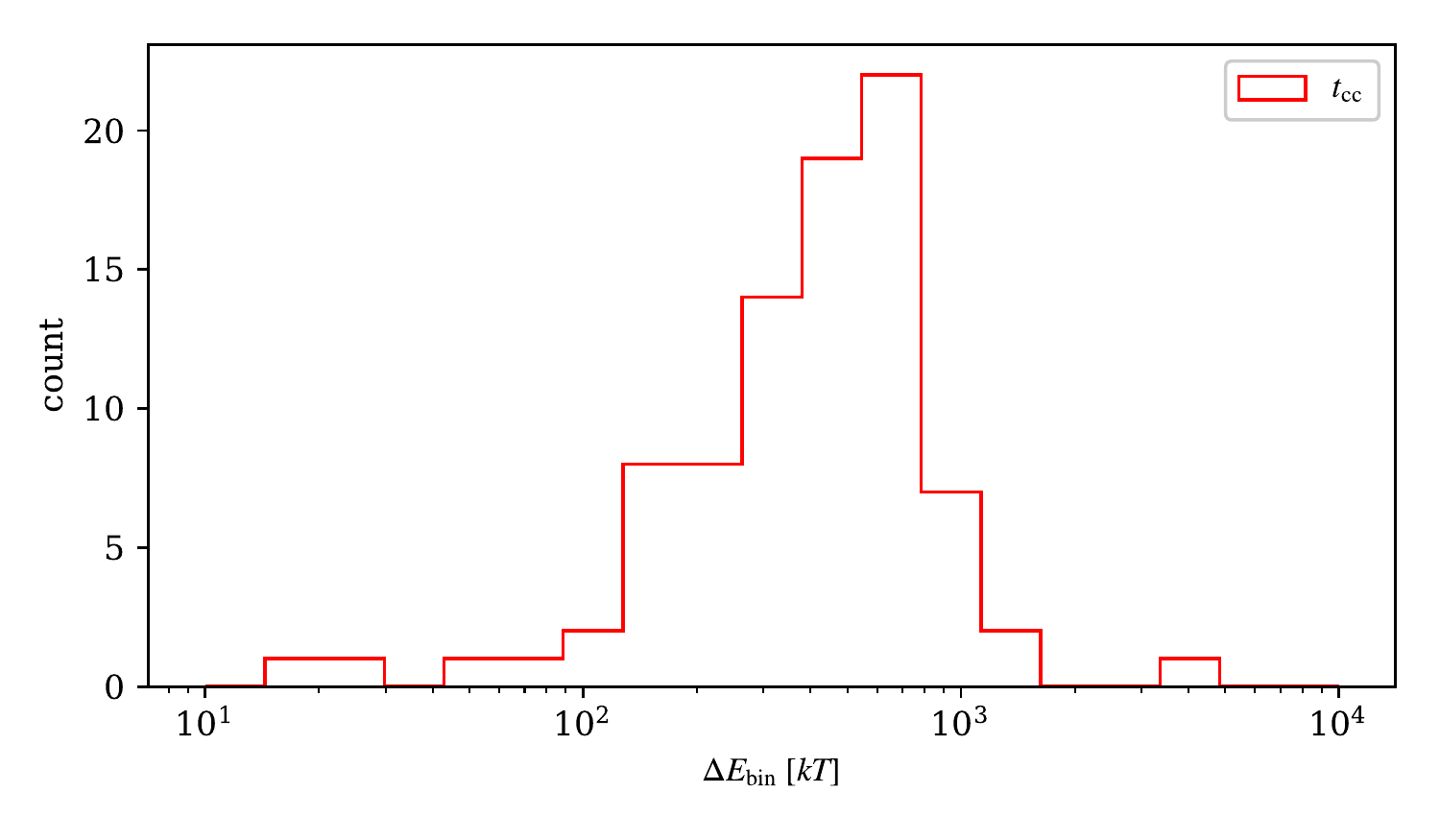}
  \caption{Distribution of the gain of binding energy, $\Delta\Ebin$, of the binary that hardened the most during the core collapse in model \emod.}
  \label{fig:E_smod}
% \end{figure}
  
  \vspace*{\floatsep}
  
% \begin{figure}
  \includegraphics[width=\linewidth]{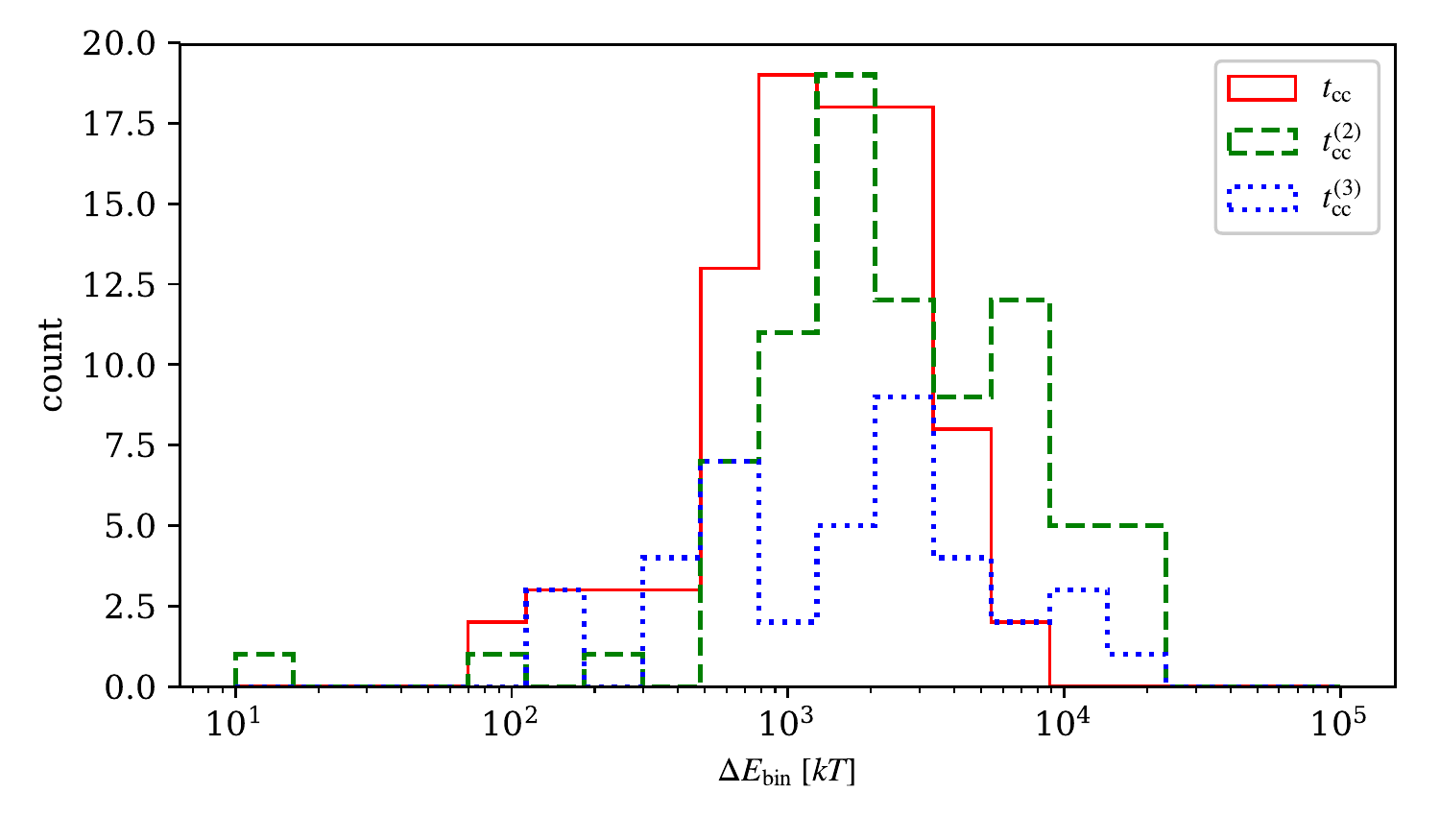}
  \caption{Distribution of the gain of binding energy, $\Delta\Ebin$, of the binary that hardened the most during a given homologous collapse in model \mmod.}
  \label{fig:E_mmod}
\end{figure}
  
\begin{figure}
  \includegraphics[width=\linewidth]{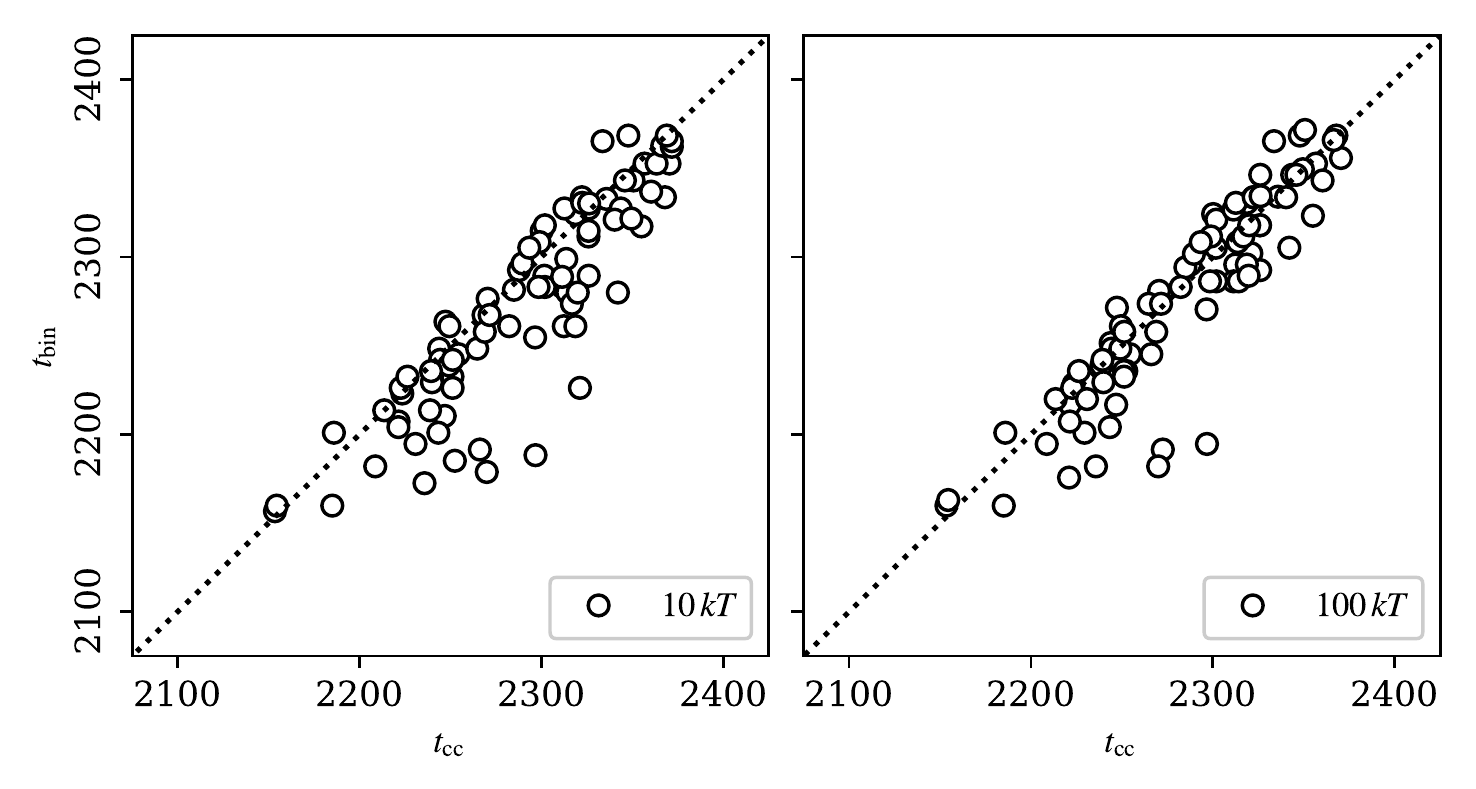}
  \caption{Time of core collapse in model \emod\ versus the time of formation of the first binary of energy $\Ebin \geq \Elim$ (indicated in each panel). The corresponding correlation coefficients are listed in Table~\ref{tab:corr}.}
  \label{fig:Tcc_Tbin_s10k}
% \end{figure}

  \vspace*{\floatsep}
  
% \begin{figure}
  \includegraphics[width=\linewidth]{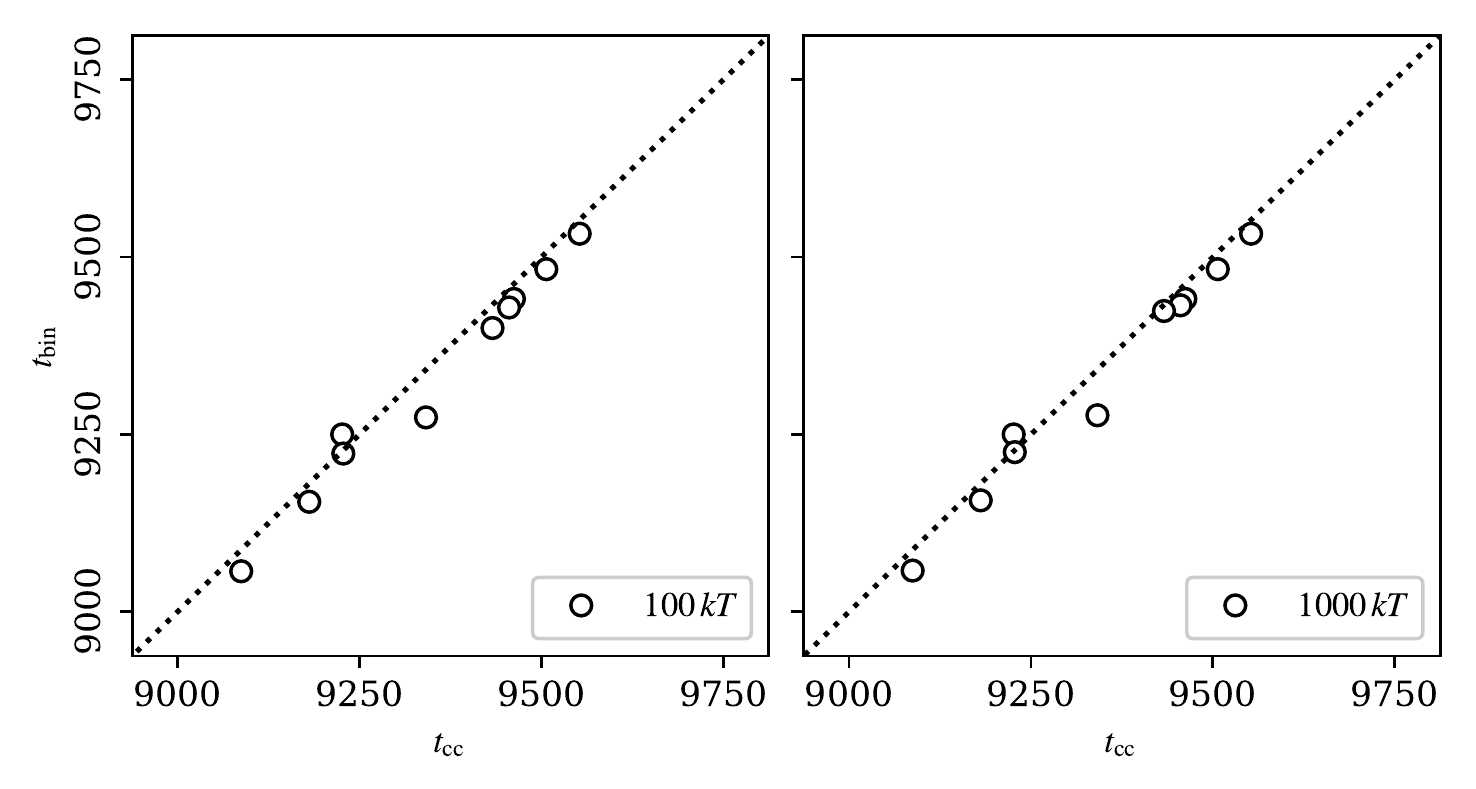}
  \caption{Same as in Fig.~\ref{fig:Tcc_Tbin_s10k} but for model \emodi.}
  \label{fig:Tcc_Tbin_s50k}
% \end{figure}
  
  \vspace*{\floatsep}
  
% \begin{figure}
  \includegraphics[width=\linewidth]{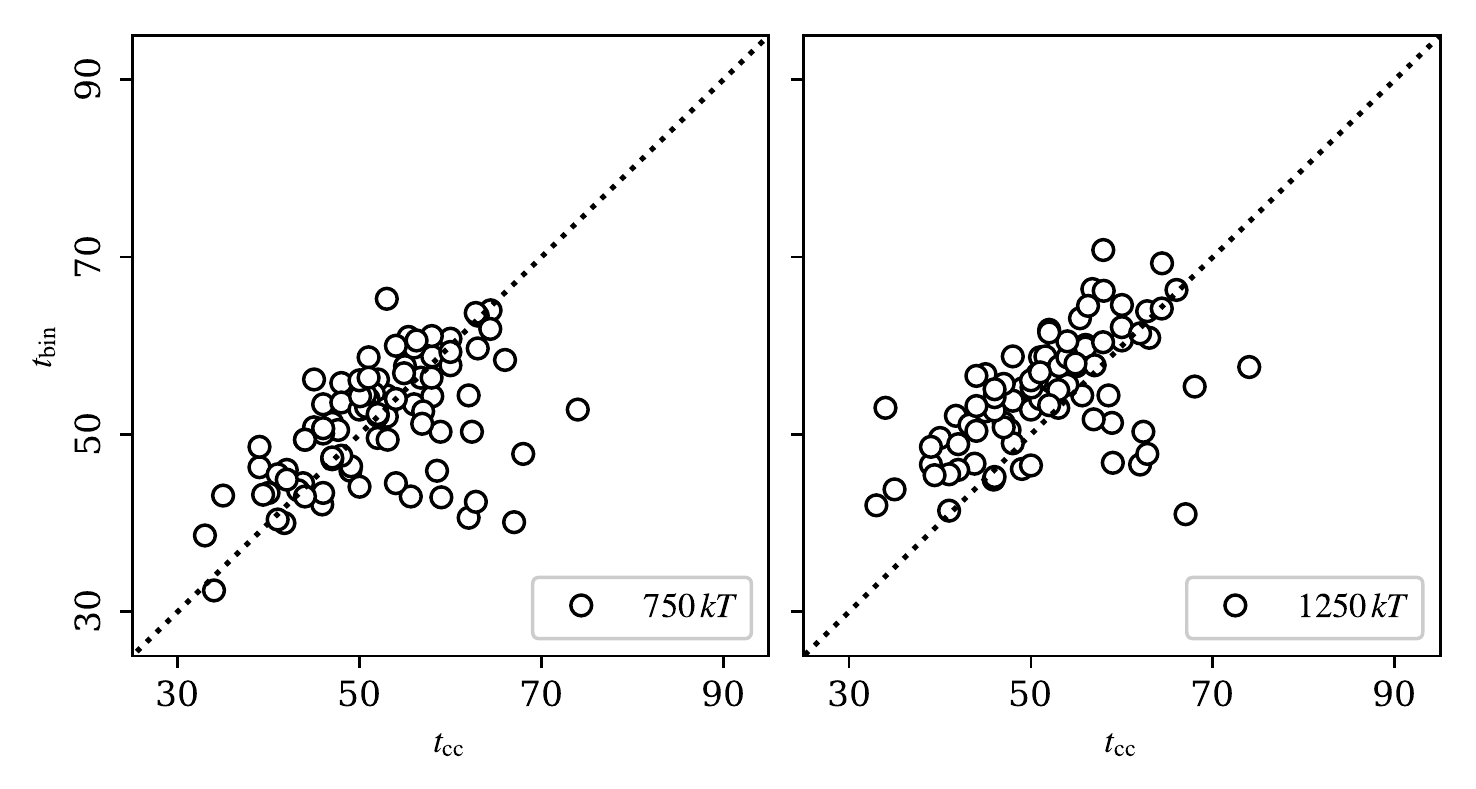}
  \caption{Same as in Fig.~\ref{fig:Tcc_Tbin_s10k} but for model \mmod. The corresponding correlation coefficients are listed in Table~\ref{tab:corr}.}
  \label{fig:Tcc_Tbin_m20k}
% \end{figure}
  
  \vspace*{\floatsep}
  
% \begin{figure}
  \includegraphics[width=\linewidth]{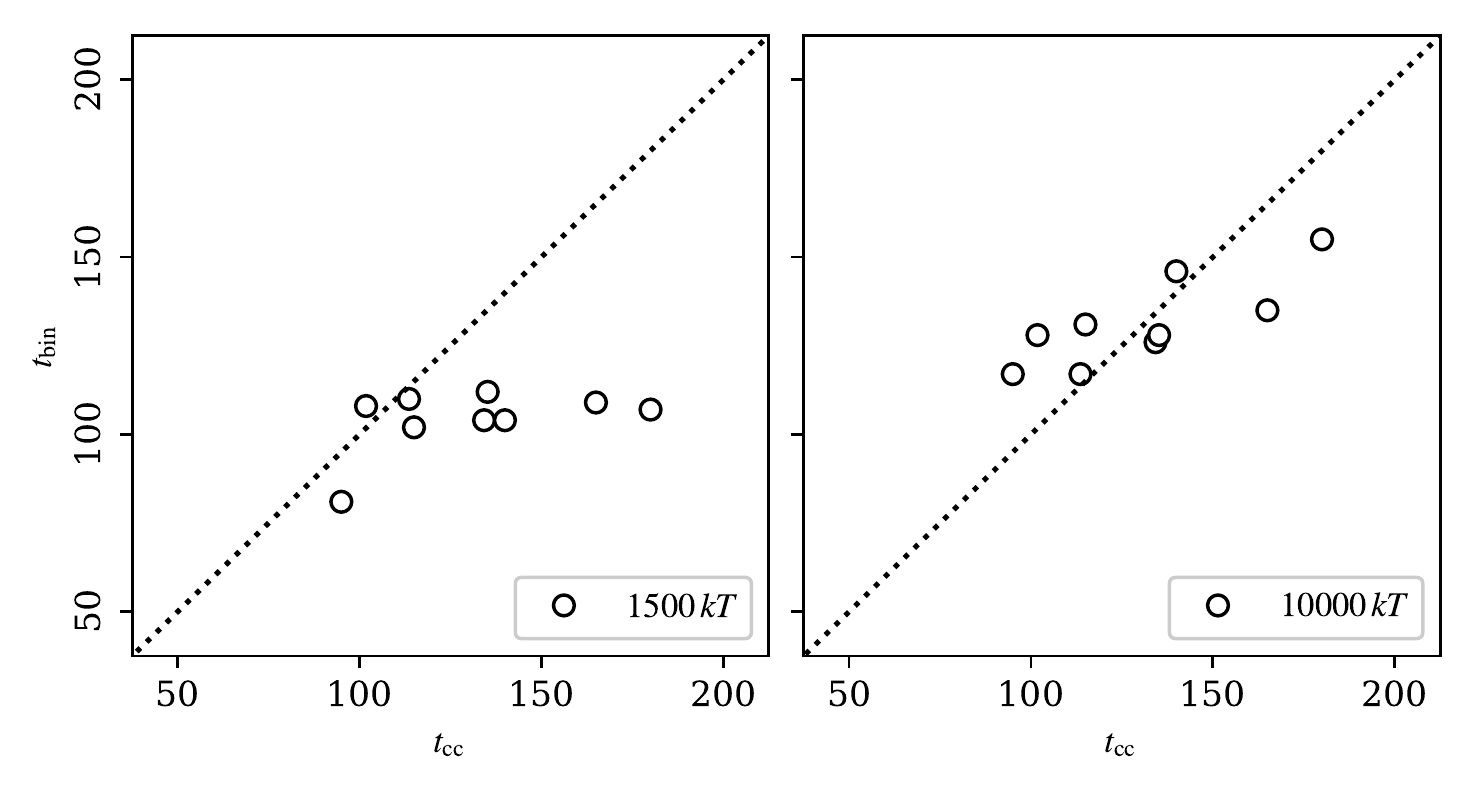}
  \caption{Same as in Fig.~\ref{fig:Tcc_Tbin_s10k} but for model \mmodi.}
  \label{fig:Tcc_Tbin_m100k}
\end{figure}

\end{document}